%% file: main.tex
\documentclass[lettersize,journal]{IEEEtran}
\usepackage{amsmath,amsfonts,amssymb}
\usepackage{algorithm}
\usepackage{array}
\usepackage{textcomp}
\usepackage{stfloats}
\usepackage{url}
\usepackage{verbatim}
\usepackage{graphicx}
\usepackage{cite}
\usepackage[caption=false,font=footnotesize]{subfig}
\usepackage{makecell}
\usepackage[dvipsnames]{xcolor} 
\usepackage{soul} 
\usepackage{tikz}
\usepackage{pgfplots} 
\pgfplotsset{compat=1.17}
\usepgfplotslibrary{colormaps}
\usepackage{multirow}
\usepackage{adjustbox}
\usepackage[colorlinks=true, linkcolor=ForestGreen, citecolor=MidnightBlue, urlcolor=blue]{hyperref} 
\usepackage{algorithmicx}
\usepackage{algpseudocode} 
\usepackage{mathtools}  
\usepackage{balance}

\DeclareMathOperator{\diag}{diag}
\newcommand{\transpose}{\mathsf{T}}

\newcommand{\re}{\mathbb{R}}
\newcommand{\pAngle}[2]{\langle#1,#2\rangle}
\newcommand{\norm}[1]{\lVert #1 \rVert}
\usepackage{accents}
\newcommand{\ubar}[1]{\underaccent{\bar}{#1}}

\DeclareSymbolFont{bbold}{U}{bbold}{m}{n}
\DeclareSymbolFontAlphabet{\mathbbold}{bbold}

\usepackage{mdframed}

\usepackage{amsthm} 

\theoremstyle{remark}
\newtheorem{remark}{Remark}

\allowdisplaybreaks

\begin{document}

\title{Optimizing DER Aggregate Flexibility via Network Reconfiguration}

\author{Feixiang~Zhang, 
        Hongyi~Li,~\IEEEmembership{Member,~IEEE},
        Bai~Cui,~\IEEEmembership{Member,~IEEE}, and
        Zhaoyu~Wang,~\IEEEmembership{Fellow,~IEEE}
\thanks{F. Zhang, H. Li, B. Cui, and Z. Wang are with the Department
of Electrical and Computer Engineering, Iowa State University, Ames,
IA, 50011 USA. Emails: \texttt{\{zfx, hongyili, baicui, wzy\}@iastate.edu}}}



\maketitle

\begin{abstract}
The aggregate flexibility region of distributed energy resources (DERs) quantifies the aggregate power shaping capabilities of DERs. It characterizes the distribution network's potential for wholesale market participation and grid service provision at the transmission level. To enhance flexibility and fully exploit the potential of DERs, this paper proposes a method to optimize the aggregate flexibility region through distribution network reconfiguration. First, we formulate the ellipsoidal aggregate flexibility region characterization problem as a two-stage adaptive robust optimization problem and derive an exact convex reformulation with a large number of second-order cone constraints. By exploiting the problem structure, we propose a scalable Benders decomposition algorithm with provable finite convergence to the optimal solution. Finally, we propose an optimal reconfiguration problem for aggregate flexibility region optimization and solve it using the custom Benders decomposition. Numerical simulations on the IEEE 123-bus test feeder demonstrate that, compared to existing approaches, substantial improvements in the aggregate flexibility region can be achieved over multiple scenarios with the optimized topology. 
\end{abstract}


\section{Introduction}
\IEEEPARstart{W}{ith} the growing emphasis on sustainable development, the large-scale integration of distributed energy resources (DER) has become increasingly important. Aggregated DERs not only reduce reliance on centralized fossil fuel generation but also enhance grid reliability. For instance, solar photovoltaic (PV) coupled with energy storage (ES) systems can help mitigate voltage fluctuations~\cite{Wang2016decentralized} and support power balance~\cite{Wang2015coordinated}. However, the inherent intermittency of DERs, especially PV and wind, introduces operational uncertainty at both the distribution and transmission levels. In addition, the flexibility of individual DERs is often limited by their sizes and geographic dispersion. As such, the effective aggregation and coordination of heterogeneous DERs become important. In this context, aggregate flexibility, which refers to the power shaping capabilities of aggregated DERs, plays an important role in market participation and ancillary service provision. 


There have been extensive research efforts on characterizing and harnessing the flexibility of standalone DERs. These studies focus on characterizing the flexibility of homogeneous DER types, such as controllable loads \cite{Zhao2017geometric}, PV systems, ES systems \cite{Alipour2022energy}, heating, ventilation, and air-conditioning (HVAC) units \cite{Tian2022real}, and electric vehicles (EVs) \cite{panda2024efficient}. Although these studies provide valuable insights into the flexibility of homogeneous DER types at the component level, no grid-level aggregation or coordination are considered. Recent works have progressed toward the quantification of the aggregate flexibility potential of heterogeneous DER ensembles, where several challenging issues need to be considered, including network constraints, load and DER uncertainties, and temporal coupling of aggregate flexibility \cite{wen2022aggregate}.

{A fundamental requirement in aggregate flexibility characterization is the feasibility of all aggregate trajectories inside the characterized region. Existing studies have addressed this issue for several polytope-based representations, including hyperbox \cite{chen2020aggregate, chen2021leveraging, Li2024distribution} and storage-like/virtual-battery models \cite{Wen2023aggregate,wen2024improved,tan2024optimal,wang2021aggregate}. Ellipsoidal representations have also been used to provide compact inner approximations of aggregate flexibility regions. The feasibility of points in the ellipsoid is guaranteed by imposing predefined affine decision rules, which may restrict the recourse decisions and lead to conservative flexibility characterization \cite{cui2021network}. Therefore, although ellipsoidal regions provide a compact temporal representation of aggregate power trajectories, certifying the feasibility of all points inside an ellipsoidal region remains challenging.}

{Existing uncertainty-aware flexibility studies can be categorized into probabilistic methods and robust optimization methods. For example, forecast errors in renewable generation and loads are modeled using Gaussian mixture models and handled via chance constraints with quantile-based deterministic reformulations \cite{wang2024stochastic}. Wasserstein distributionally robust joint chance constraints are used to derive reliable aggregate feasible regions when the true uncertainty distribution is unknown \cite{Zhou2025aggregated, wen2024centralized}. In addition, Chebyshev-based chance constraints are adopted in~\cite{wang2025multi} to construct uncertainty-aware inner approximations using moment information of forecast errors. These probabilistic methods improve reliability under uncertainty, but they do not guarantee robustness against all uncertainty realizations. In contrast, a robust ellipsoidal flexibility characterization problem is formulated with a norm-bounded load uncertainty set in~\cite{cui2021network}. However, it adopts predefined recourse decision rules to maintain tractability. This indicates that robust feasibility certification under norm-bounded uncertainty remains insufficiently addressed.}

Operational reconfiguration is effective in managing distribution systems with high DER penetrations, and its capability to provide economic, reliability, and resilience benefits has been extensively analyzed \cite{dorostkar2016value}. However, most existing works on aggregate flexibility assessment assume fixed distribution network topology. Since the aggregate flexibility region is constrained not only by DER characteristic but also by network constraints, changes in network topology directly impact the flexibility region and could unlock significant flexibility benefits for energy and grid service provision. Without modeling switching capability explicitly, flexibility may be systematically underestimated and under-utilized, especially in rural areas with long and lightly loaded feeders or for highly congested ones in urban settings. {This motivates us to consider network reconfiguration that maximizes the aggregate flexibility region of DERs. In the proposed formulation, network topology and aggregate flexibility region are jointly optimized.}

However, the introduction of binary variables {that model the network topologies} complicates the problem since i) the flexibility characterization needs to be adaptive to the variable topology, and ii) the resulting problem becomes a mixed-integer optimization problem with significantly higher computational complexity. Due to the technical difficulties, leveraging reconfiguration strategies for aggregate flexibility improvement remains under-explored \cite{li2025aggregate} despite the potential improvement. For example, \cite{churkin2023impacts} compares aggregate flexibility by enumerating all candidate topologies. However, this approach is not scalable to realistically sized systems and it does not account for intertemporal coupling of DER operations.

{Therefore, in this paper, we propose a unified and computationally tractable framework to perform distribution reconfiguration that optimizes the ellipsoidal aggregate flexibility region, while jointly modeling intertemporal DER operations and ensuring robustness against norm-bounded load uncertainties. These components are integrated within a single ARO structure so that a rigorous assessment of the joint impact of topology, DER characteristics, and uncertainties on the aggregate flexibility can be made.} The main contributions of this paper are threefold:

{First, we introduce the first unified optimization framework that optimizes the aggregate flexibility region through network reconfiguration while capturing intertemporal coupling among DERs and load uncertainties. This contrasts with prior works on aggregate flexibility characterization that either assume a fixed topology, use conservative approximations, ignore intertemporal coupling, or assume a deterministic model.}

{Second, we derive an exact reformulation of the ARO problem for ellipsoidal aggregate flexibility characterization. The ARO problem can guarantee that every point inside the ellipsoid region admits feasible second-stage DER disaggregation decisions under all uncertainty realizations within the norm-bounded uncertainty set. By using duality and support functions, the semi-infinite robust feasibility condition is converted into a finite set of tractable cuts. Compared to our earlier ellipsoidal characterization \cite{cui2021network}, the proposed approach provides a provably stronger characterization of the aggregate flexibility region.}

{Third, we design a customized Benders decomposition algorithm that handles binary variables and robust feasibility constraints. We also develop an efficient reformulation that significantly reduces the number of non-convex bilinear terms and improves the global solver's performance in solving the subproblem. This enables the scalability of the proposed ARO for the reconfiguration of large-scale feeders. Numerical experiments on the IEEE 123-bus test feeder demonstrate that the reconfiguration improves the aggregate flexibility region
by more than 100\%, especially under significant uncertainties.}


The remainder of the paper is organized as follows: Section \ref{sect:background} presents the background, including notations, DER and system modeling, and basic optimization models. Section \ref{sect:technical} develops the ARO formulation and its solution algorithm for the aggregate flexibility characterization under a fixed topology. Section \ref{sect:reconfig} extends the formulation and solution algorithm to consider network reconfiguration. Numerical simulations that demonstrate the effectiveness of the proposed approach are presented in Section \ref{sect:simulation}. Finally, Section \ref{sect:conclusion} concludes the paper and summarizes the main findings.

\section{Background} \label{sect:background}
\subsection{Network Model and Notation}
We model the distribution system (network) as a simple, connected, undirected graph $(\mathcal{N}, \mathcal{E})$ where $\mathcal{N} = \{ 0, 1, \ldots, N\}$ is the set of buses and $\mathcal{E} \subseteq \mathcal{N} \times \mathcal{N}$ is the set of distribution lines. Let $\mathcal{N}_1 = \{ 1, 2, \ldots, N\}$ denote the set of load buses and bus $0$ be the substation (slack) bus, so that $\mathcal{N} = \mathcal{N}_1 \cup \{0\}$. A feasible network topology is a spanning tree (radial subgraph) of $(\mathcal{N}, \mathcal{E})$. Each distribution line $(i,j) \in \mathcal{E}$ has resistance $r_{ij}$ and reactance $x_{ij}$.

We denote by $\mathcal{E}_s \subseteq \mathcal{E}$ the subset of lines equipped with controllable switches. For each $(i,j) \in \mathcal{E}_s$, a binary variable $s_{ij}$ indicates whether the line is closed ($s_{ij} = 1$) or open ($s_{ij} = 0$). For notational uniformity, we also define $s_{ij}=1$ for all non-switchable lines $(i,j) \in \mathcal{E}\setminus\mathcal{E}_s$. The operating topology is the subgraph formed by all non-switchable lines and the closed switchable lines.

We consider aggregate flexibility over a multi-period horizon, where each time step $\tau$ may represent $5$-, $15$-, or $60$-minute interval depending on the application. The time horizon is indexed by $\mathcal{T} = \{1, 2, \ldots, T\}$. For notational convenience, we also define $\mathcal{T}_2 = \{2, 3, \ldots, T\}$ as the time horizon excluding the initial period. For each bus $i \in \mathcal{N}$ and time $t \in \mathcal{T}$, $p_{i,t}$ ($q_{i,t}$) denotes the real (reactive) power injection and $v_{i,t}$ the squared voltage magnitude. For line $(i,j) \in \mathcal{E}$ and time $t \in \mathcal{T}$, $p_{ij,t}$ ($q_{ij,t}$) denotes the real (reactive) power flow from bus $i$ to bus $j$ at time $t$. 

\textit{Notations:} An all-one vector of appropriate dimension is denoted by $\mathbbold{1}$. $\diag(\cdot)$ denotes a diagonal matrix with its argument on the main diagonal. \(\langle x, y \rangle := x^\top y\) denotes the standard inner product.
\(\|x\|_2 := \sqrt{x^\transpose x}\) denotes the Euclidean ($\ell_2$) norm. For a positive semidefinite matrix $Q \in \mathbb{R}^{n\times n}$, the $Q$-weighted vector-norm is
\begin{equation}
    \norm{v}_Q = \norm{Q^{\frac{1}{2}}v}_2, \qquad v \in \mathbb{R}^n,
\end{equation}
where $Q^{1/2}$ is the principal square root of $Q$.


\subsection{DER Models} \label{sect:DER}

We represent each controllable DER by an index $d\in\mathcal{D}$, where $\mathcal{D}$ is the set of all controllable DERs in the system. Let $i(d)$ denote the bus at which DER $d$ is connected. The real power output of DER $d$ at time $t \in \mathcal{T}$ is denoted by $x_{d,t}$. For simplicity, we assume that each DER operates at a constant power factor. Other inverter-based control modes (e.g., volt-var control or $PQ$-capability curve) can be incorporated without affecting the overall formulation.

Different DER types impose different operational constraints, which we represent through technology-specific feasible sets. For each DER $d$, its real power trajectory over the time horizon $\mathcal{T}$ satisfies: 
\begin{equation}
    x_d := (x_{d,t})_{t\in\mathcal{T}} \in \mathcal{X}_d,
\end{equation}
where $\mathcal{X}_d$ is a polytope. For illustration, the feasible set of a PV system $d$ is defined as
\begin{equation}
    \mathcal{X}_d = \bigl\{ x_{d} \in \mathbb{R}^T \mid 0 \le x_{d,t} \le \bar{p}_{d,t}, \, t\in\mathcal{T} \bigr\},
\end{equation}
where $\bar{p}_{d,t}$ is the available PV generation at time $t$. Detailed definitions of the feasible sets $\mathcal{X}_d$ for other DER types are provided in Appendix \ref{app:DER}. The overall DER feasible set is therefore the Cartesian product:
\begin{equation} \label{eq:constraint:DER}
    \mathcal{X} := \prod_{d\in\mathcal{D}} \mathcal{X}_d.
\end{equation}


\subsection{Uncertainty Model: Uncontrollable Load} \label{sect:uncertainty}
In addition to the controllable DERs, we model the uncontrollable real power load at each bus $k \in \mathcal{N}_1$ and time $t\in\mathcal{T}$ as a multiplicative perturbation of its nominal value $p_{k,t}^{u,0}$:
\begin{equation}
    p_{k,t}^u = p_{k,t}^{u,0} (1 + \delta \zeta_{c(k),t}), 
\end{equation}
where $c(k) \in \{\mathrm{res, com, ind}\}$ denotes the load category (residential, commercial, and industrial) at bus $k$; $\delta > 0$ is the uncertainty level; and $\zeta_{c,t}$ captures the normalized load deviation for load category $c$ at time $t$. We assume constant power factor for the uncontrollable loads.

To model load uncertainty correlations among load categories, we collect the normalized load deviations for different load categories into a vector:
\begin{equation}
    \zeta_t = \begin{bmatrix}
        \zeta_{\mathrm{res}, t} \\
        \zeta_{\mathrm{com}, t} \\
        \zeta_{\mathrm{ind}, t}
    \end{bmatrix}.
\end{equation}
We assume $\zeta_t$ lies in an $\ell_2$-norm-bounded uncertainty set:
\begin{equation}
    \zeta_t \in \mathcal{U} = \bigl\{ \zeta_t \in \mathbb{R}^3 \mid \|\zeta_t\|_2 \le 1 \bigr\}.
\end{equation}
It follows that loads in different categories lie within an ellipsoidal uncertainty set. The uncertainty set captures joint deviations across load categories and the correlation within the same load category. {It should be noted that this paper focuses on norm-bounded uncertainty sets as a tractable representation of the joint bounded deviations of different load categories. The proposed framework is not limited to this specific uncertainty set, and other uncertainty models, such as budget uncertainty sets, can also be incorporated without major changes to the proposed framework.}


It follows from the preceding models that the net real power injection at load bus $i \in \mathcal{N}_1$ and time $t$ can be represented as
\begin{equation}
    p_{i,t} = \sum\nolimits_{d\in\mathcal{D}: i(d) = i} x_{d,t} - p_{i,t}^u(\zeta_t).
\end{equation}
The reactive power injection can be represented similarly.


\subsection{Linear Distribution Power Flow Model} \label{sect:LinDistFlow}
We adopt the LinDistFlow model \cite{Baran32} to approximate power flows in the radial distribution system. This linearized model provides accurate affine relationships between bus power injections and squared voltage magnitudes under nominal operating conditions. It is widely used in distribution system optimization and reconfiguration studies. 

We assume the network is radial and assign an orientation to each distribution line $(i,j) \in \mathcal{E}$ by directing all the lines away from the substation bus. Thus, every line is oriented from the upstream bus to the downstream bus along the unique path from the substation bus. The real and reactive power flows $p_{ij, t}$ and $q_{ij,t}$ are defined with respect to this orientation. For notational simplicity, we omit the time index $t$ in the LinDistFlow model, which can be defined as:
\begin{subequations} \label{eq:LinDistFlow}
\begin{align}
    & p_i = \sum\nolimits_{(i,j) \in \mathcal{E}} p_{ij} - \sum\nolimits_{(k,i) \in \mathcal{E}} p_{ki}, && \forall i \in \mathcal{N}_1, \label{eq:LinDistFlow:p} \\
    & q_i = \sum\nolimits_{(i,j) \in \mathcal{E}} q_{ij} - \sum\nolimits_{(k,i) \in \mathcal{E}} q_{ki}, && \forall i \in \mathcal{N}_1, \label{eq:LinDistFlow:q} \\
    & v_j = v_i - 2\bigl( r_{ij} p_{ij} + x_{ij}q_{ij} \bigr), && \forall (i,j)\in\mathcal{E} \label{eq:LinDistFlow:v}
\end{align}
\end{subequations}

The real power injection $p_0$ at the substation bus (the power imported from the upstream transmission system) is given by
\begin{equation}
    p_0 = -\sum\nolimits_{i: (0,i) \in \mathcal{E}} p_i.
\end{equation} 

The squared voltages satisfy
\begin{equation} \label{eq:voltage}
    \ubar{v}_i \le v_i \le \bar{v}_i, \qquad i \in \mathcal{N},
\end{equation}
with the substation voltage fixed (\(\ubar{v}_0 = \bar{v}_0\)).

\subsection{Reconfiguration Formulation}

To enable flexibility improvements through reconfiguration, explicit modeling of switch statuses and the network radiality constraints are needed. Several widely used radiality formulations exist, including cycle elimination \cite{borghetti2012mixed}, path-based \cite{ramos2005path}, virtual flow \cite{lavorato2012imposing, singh2022joint}, and parent-child models \cite{taylor2012convex}. In this work, we adopt the single-commodity flow (SCF) formulation, which can be interpreted as an extended formulation of the cutset description of the spanning tree polytope. The SCF formulation provides a good balance between tightness and model complexity.


Under this formulation, every load bus has one unit of fictitious demand $\ell_i$, and the substation bus is the only source of the fictitious flow. We adopt the convention that positive $\ell_i$ indicates injection. Adopting the same line orientation convention from Section \ref{sect:LinDistFlow}, the SCF formulation is 
\begin{subequations} \label{eq:SCF}
\begin{align}
    & \ell_0 = N, && \\
    & \ell_i = -1, && i \in \mathcal{N}_1, \\
    & \ell_i = \sum\nolimits_{(i,j) \in \mathcal{E}} \ell_{ij} - \sum\nolimits_{(k,i) \in \mathcal{E}} \ell_{ki}, && i \in \mathcal{N}, \label{eq:SCF:balance}\\
    & -s_{ij}N \le \ell_{ij} \le s_{ij}N, && (i,j) \in \mathcal{E}, \\
    & \sum\nolimits_{(i,j) \in \mathcal{E}} s_{ij} = N, \label{eq:SCF:cardinality}
\end{align}
\end{subequations}
where $\ell_{ij}$ is the flow of the fictitious load on line $(i,j)\in \mathcal{E}$. The flow balance constraints \eqref{eq:SCF:balance} enforce that every load bus must be connected to the substation, and the cardinality constraint \eqref{eq:SCF:cardinality} ensures that the resulting connected topology has exactly $N$ closed lines to form a spanning tree. 

To account for topology changes, the following big-$M$-based constraints are added to the LinDistFlow model \eqref{eq:LinDistFlow}:
\begin{equation}
    -s_{ij} M \leq (p_{ij}, q_{ij}) \leq s_{ij} M, \quad (i,j) \in \mathcal{E}.
\end{equation}
In addition, \eqref{eq:LinDistFlow:v} is modified as
\begin{align} \label{eq:LinDistFlow:reconfig:v}
    & v_i - 2\bigl( r_{ij} p_{ij} + x_{ij}q_{ij} \bigr) - M(1-s_{ij}) \le v_j \nonumber\\
    & \hspace{0.1in} \le v_i - 2\bigl( r_{ij} p_{ij} + x_{ij}q_{ij} \bigr) + M(1-s_{ij}), \; (i,j) \in \mathcal{E}
\end{align}

\subsection{Compact Representation of Overall Constraints}

All DER, network, voltage, and uncertainty constraints \eqref{eq:constraint:DER}--\eqref{eq:LinDistFlow:reconfig:v} can be written in the following compact form:
\begin{subequations}
\begin{align}
    & Ax \ge d - Bp_0 - D\zeta - C s, \label{eq:compact:Ax} \\
    & \zeta \in \mathcal{U}^T = \bigl\{ \zeta \in \mathbb{R}^{3T} \mid \|\zeta_t\|_2 \le 1, \,t \in \mathcal{T} \bigr\}. \label{eq:compact:uncertainty}
\end{align}
\end{subequations}
where, for notational simplicity, we slightly abuse notation and overload $x$ as the concatenation of all operational variables, including DER power trajectories, bus voltages, line flows, and any auxiliary device-level state variables.

\input{section_technical_1.tex}

\input{section_reconfig.tex}
\input{section_numerical_simulations.tex}

\section{Conclusion} \label{sect:conclusion}
This paper proposes a method to optimize the aggregate flexibility region of DERs through network reconfiguration. The optimal reconfiguration problem is formulated as a three-stage ARO problem, which is subsequently reformulated as a large-scale MISDP problem and solved by a custom scalable Benders decomposition algorithm. The efficiency and effectiveness of the proposed approach are validated using the IEEE 123-bus test feeder. Additionally, the influence of uncertainty levels and computing time on the flexibility region is analyzed.

The simulation results show the improvement in flexibility region characterization by the proposed method. They also show network reconfiguration improves the aggregate flexibility region by over 91.56\% compared to the reference topology. The proposed approach effectively alleviates constraint violations by altering the network topology. The results also suggest the benefit of network reconfiguration in improving the aggregate flexibility region as the uncertainty increases.

{Several directions remain for future work. On the modeling side, the current framework adopts the LinDistFlow approximation and a constant power factor assumption to preserve tractability. Future work will investigate more accurate power flow models, data-driven feasibility verification methods, and more general DER control modes, such as Volt-Var, Watt-Var, constant reactive power, Volt-Watt control, and P-Q capability regions. In addition, switching costs, coordination costs, operational logistics, and more data-driven uncertainty set construction methods will be incorporated to improve the practical applicability of the proposed framework.}

\input{section_appendices.tex}


\balance
\bibliographystyle{IEEEtran}
\bibliography{IEEE_Xplore_Citation_BibTeX}

\vfill

\end{document}

%% file: section_technical_1.tex
\section{Characterization of Ellipsoidal Aggregate Flexibility Region}\label{sect:technical}

The aggregate flexibility region characterization is a key building block in the proposed reconfiguration framework. Before presenting the formulation and solution method for the optimal reconfiguration problem, we introduce a new algorithm for this key building block. The algorithm is based on Benders decomposition and differs from our previous policy-based approximation method~\cite{cui2021network}. 
By exploiting the problem structure, it guarantees finite convergence to the optimal solution, in contrast to the earlier method that provides only an approximate solution with a bounded tightness factor.

\subsection{Definition and Motivation}
The substation power injection across the time horizon $p_0 := (p_{0,t})_{t\in\mathcal{T}}$ represents the trajectory of the net real power exchanged with the transmission system. Characterizing the set of feasible trajectories $p_0$ is challenging due to i) network constraints; ii) DER constraints, many of which introduce intertemporal coupling (HVAC thermal dynamics, ES SoC evolution, EV cumulative energy limits); and iii) uncertainty in uncontrollable loads. 

The \emph{aggregate flexibility region} is defined as the set of all power trajectories $p_0$ for which a feasible second-stage variable $x$ exists under all uncertainty realizations:
\begin{align} \label{eq:AFregion}
\mathcal{P} = \bigl\{ p_0 \in \mathbb{R}^T \mid \forall \zeta \in \mathcal{U}^T,\, & \exists\, x(\zeta): \nonumber\\
& Ax(\zeta) \ge d-Bp_0 - D\zeta \bigr\},
\end{align}
Notice that the $Cs$ term is omitted since we only consider fixed topologies in this section.

\begin{remark}[Convexity and Hardness of Exact Characterization]
    The aggregate flexibility region $\mathcal{P}$ in \eqref{eq:AFregion} is convex. To see this, notice that for any fixed uncertainty realization $\zeta$, the corresponding feasible set $\{\, p_0 \mid \exists\, x: Ax \ge d - Bp_0 - D\zeta \,\}$ is convex since it is the projection of a polyhedral set in the $(p_0, x)$-space onto the $p_0$-coordinates. Since $\mathcal{P}$ is the intersection of these convex sets over all $\zeta \in \mathcal{U}^T$, the overall region is convex. 

    Despite the convexity, explicitly characterizing $\mathcal{P}$ is computationally intractable. Even computing the exact maximum-volume ellipsoid inscribed in $\mathcal{P}$ amounts to projection of high-dimensional polyhedral, which is known to be NP-hard (see, e.g., \cite{cui2021network} and references therein). By exploiting problem structure and using Benders decomposition with global optimization in the subproblem, we detail in the remainder of the section how we obtain a scalable solution method for finding the max-volume inscribing ellipsoid that remains efficient for large distribution networks.\qed
\end{remark}

\subsection{Problem Formulation}
We denote by $\mathcal{E}(Q,c) := \bigl\{ x\in\re^n \mid (x-c)^\top Q^{-1}(x-c) \le 1 \bigr\}$ an ellipsoid centered at $c \in \re^n$ with $Q \succeq 0$. 
The ellipsoidal aggregate flexibility region characterization problem can be formulated as the following robust feasibility problem:
\begin{subequations} \label{eq:ARO_semiinfty}
\begin{align}
    \max_{Q \succeq 0, c} \quad & \log\det Q \label{eq:ARO_semiinfty:a}\\
    \text{s.t.} \quad & \max_{\xi \in \mathcal{E}(Q,c),\zeta\in\mathcal{U}^T} \min_{x \in \mathcal{W}(\xi,\zeta)} g^\top x \le 0. \label{eq:ARO_semiinfty:b} 
\end{align}
\end{subequations}
The problem seeks to find the maximum-volume ellipsoid satisfying robust feasibility constraints since maximizing $\log\det Q$ maximizes the ellipsoid's volume \cite[Sec. 8.3.1]{boyd2004convex}. Here, the ellipsoid $\mathcal{E}(Q,c)$ is an inner approximation of the aggregate flexibility region $\mathcal{P}$. {The proposed framework can also accommodate other tractable objectives, such as weighted flexibility or cost-aware objectives with linear cost terms.}

The constraint set for the second-stage variable $x$ is
\begin{equation}
    \mathcal{W}(\xi,\zeta) := \{ x \in \re^m \mid Ax \ge d - B\xi -D\zeta\}. 
\end{equation}
The inner minimization problem in \eqref{eq:ARO_semiinfty:b} is a feasibility check problem: it checks whether there exists a recourse $x(\zeta) \in \mathcal{W}(\xi,\zeta)$ for a given flexibility trajectory $\xi$ and worst-case uncertainty $\zeta$ such that the cost $g^\top x(\zeta) \le 0$. If such a recourse exists for all $\zeta \in \mathcal{U}^T$, then $\xi$ is feasible ($\xi\in\mathcal{P}$). Otherwise, there is an uncertainty realization $\zeta$ such that $g^\top x > 0$ for all $x \in \mathcal{W}(\xi,\zeta)$, and $\xi \notin \mathcal{P}$. 

We introduce nonnegative slack variables in the recourse feasible set $\mathcal{W}(\xi,\zeta)$ so that every operational constraint is relaxed by a slack variable. The objective function minimizes the sum of these slack variables. This construction ensures complete recourse: the feasibility check problem is feasible for all $(\xi,\zeta)$. For brevity, we use the same notation as \eqref{eq:compact:Ax} to denote the augmented recourse vector and constraint set.


At first glance, model \eqref{eq:ARO_semiinfty} appears to be a semi-infinite problem because the robust feasibility constraint \eqref{eq:ARO_semiinfty:b} is equivalent to the following set of infinitely many constraints:
\begin{subequations} \label{eq:robfeas_semiinfty}
\begin{align}
    g^\top x^e & \le 0, \\
    Ax^e & \ge d - B\xi^e - D\zeta^e, \quad \forall (\xi^e, \zeta^e) \in \mathcal{E}(Q,c)\times\mathcal{U}^T  
\end{align}
\end{subequations}
In what follows, we derive an exact finite-dimensional reformulation of \eqref{eq:robfeas_semiinfty}, replacing the infinite set above with a finite collection of constraints amenable to Benders decomposition.

\subsection{Finite-Dimensional Reformulation} \label{sect:technical:reform}
We define the cost-to-go function $C(\cdot)$ as the optimal value of the inner minimization problem in \eqref{eq:ARO_semiinfty:b} for given $(\xi, \zeta)$:
\begin{subequations}
\begin{align}
    C(\xi,\zeta) =  \min_{x} \quad & g^\top x \\
    \text{s.t.} \quad & Ax \ge d - B\xi - D\zeta
\end{align}
\end{subequations}
Since the problem has complete recourse, strong duality holds for all $(\xi,\zeta)$ pair, so
\begin{subequations}
\begin{align}
    C(\xi,\zeta) = \max_{y \ge 0} \quad & (d-B\xi-D\zeta)^\top y \\
    \text{s.t.} \quad & A^\top y = g
\end{align}
\end{subequations}
Notice the above dual problem is in standard form, so its feasible region $\mathcal{Y} := \{ y \ge 0 \mid A^\top y = g \}$ contains at least one extreme point \cite[Col. 2.2]{bertsimas1997introduction}. Let the extreme points be $y^1, \ldots, y^K$, then 
\begin{equation}
    C(\xi,\zeta) = \max_{i = 1, \ldots, K} \; (d-B\xi-D\zeta)^\top y^i.
\end{equation}

Let the value function $V(Q,c)$ be the optimal value of the $\max$-$\min$ problem in \eqref{eq:ARO_semiinfty:b}, it follows that
\begin{align}
    V(Q,c) &= \max_{\xi \in \mathcal{E}(Q,c),\zeta \in \mathcal{U}^T} C(\xi,\zeta) \nonumber\\
    &= \max_{\xi \in \mathcal{E}(Q,c),\zeta \in \mathcal{U}^T} \max_{i = 1, \ldots, K} \; (d-B\xi-D\zeta)^\top y^i.
\end{align}
We denote the support function of a set $\mathcal{C}$ by $S_\mathcal{C}(y)$, that is, $S_\mathcal{C}(y) := \sup\, \bigl\{ y^\top x \mid x \in \mathcal{C} \bigr\}$. Since the constraint sets $\mathcal{Y}$ and $\mathcal{E}(Q,c) \times \mathcal{U}^T$ are independent, the two maximization problems are interchangeable. Hence $V(Q,c)$ equals the maximum of $\pAngle{d}{\cdot} + S_{\mathcal{E}(Q,c)}(\pAngle{-B}{\cdot}) + S_{\mathcal{U}^T}(\pAngle{-D}{\cdot})$ over the $K$ extreme points because
\begin{align} \label{eq:value_doublemax}
& V(Q,c) = \max_{i=1,\ldots,K}\ \max_{\xi\in\mathcal{E}(Q,c), \zeta\in\mathcal{U}^T}
(d-B\xi-D\zeta)^\top y^i \nonumber\\
&\hspace{-0.1in} = \max_{i=1,\ldots,K} \bigl\{ d^\top y^i + S_{\mathcal{E}(Q,c)} ( -B^\top y^i )  + S_{\mathcal{U}^T} ( -D^\top y^i ) \bigr\}. 
\end{align}
To further simplify $V(Q,c)$, we note that the support functions of both the ellipsoid and the Cartesian product of Euclidean unit balls can be represented in the following closed form 
\begin{subequations}
\begin{align}
    & S_{\mathcal{E}(Q,c)} (z) = z^\top c + \norm{z}_Q, \\
    & S_{\mathcal{U}^T} (z) = \sum\nolimits_{t=1}^T \norm{z_{[t]}}_2,
\end{align}
\end{subequations}
where $z = (z_{[1]},\ldots,z_{[T]})$ with each block $z_{[t]} \in \mathbb{R}^3$. It follows
\begin{align}
    V(Q,c) =  \max_{i=1,\ldots, K} \Bigl\{ (d-Bc)^\top y^i & + \norm{B^\top y^i}_Q \nonumber\\
    & \hspace{-0.3in} + \sum\nolimits_{t=1}^T \norm{(D^\top y^i)_{[t]}}_2 \Bigr\}.
\end{align}

Since $K$ is finite, the robust feasibility constraint \eqref{eq:ARO_semiinfty:b} is equivalent to 
\begin{multline} \label{eq:bcut}
    (d-Bc)^\top y^i + \norm{B^\top y^i}_Q + \sum\nolimits_{t=1}^T \norm{(D^\top y^i)_{[t]}}_2 \le 0, \\
    i = 1, \ldots, K,
\end{multline}
which are convex in $(Q,c)$ since $\norm{B^\top y^i}_Q$ is second-order cone (SOC)-representable as $\|Q^{1/2}B^\top y^i\|_2$. Therefore, \eqref{eq:ARO_semiinfty} becomes a convex determinant-maximization problem with a finite (but large) number of SOC cuts. Instead of adding all cuts at once, we generate them on-the-fly using Benders decomposition.

\subsection{Solution Algorithm: Benders Decomposition} \label{sect:Benders}

Benders decomposition is a classic outer approximation algorithm that partitions a large problem into a master problem and a subproblem. In our case, the master problem solves for the ellipsoid parameters $(Q,c)$ with an incrementally added subset of cuts \eqref{eq:bcut} that shrinks the feasible region from the outside, while the subproblem checks robust feasibility of the candidate ellipsoid by examining its support function over the dual polyhedron. The details of the algorithm are outlined below, along with some remarks on practical implementation.

\subsubsection{Master Problem} The master problem in the $k^\text{th}$ ($k \le K$) iteration is
\begin{subequations} \label{eq:AFR_master}
\begin{align}
    \max_{Q\succeq 0, c} \quad & \log\det Q \\
    \text{s.t.} \quad & (d-Bc)^\top y^i + \norm{B^\top y^i}_Q + \sum_{t=1}^T \|(D^\top y^i)_{[t]}\|_2\le 0, \nonumber\\
    & \hspace{1.5in} \forall i \in \mathcal{I}^{(k-1)} \label{eq:AFR_master_ineqa}
\end{align}
\end{subequations}
where $\mathcal{I}^{(k-1)} := \{1, \ldots, k-1\} $ is the active index set and $y^i, i\in \mathcal{I}^{(k-1)}$ are the added extreme points of $\mathcal{Y}$. Note the master problem grows by one SOC constraint per iteration. It returns the optimal $(Q^{(k)}, c^{(k)})$ which will subsequently be checked by the subproblem.

\subsubsection{Subproblem} The subproblem employs a separation oracle to examine the feasibility of the candidate ellipsoid and, when the infeasibility is certified, strengthens the master problem by adding a feasibility cut. Specifically, given $(Q^{(k)}, c^{(k)})$ from the master problem, the separation oracle certifies the feasibility of the candidate ellipsoid by verifying the non-positivity of the optimal value $v^{(k)}$ to the following problem:
\begin{equation} \label{eq:AFR_subproblem}
    \max_{y \in \mathcal{Y}}\; (d-Bc^{(k)})^\top y + \norm{B^\top y}_{Q^{(k)}}  + \sum_{t=1}^T \|(D^\top y)_{[t]}\|_2
\end{equation}
If the optimal value $v^{(k)}$ is non-positive, the feasibility of the candidate ellipsoid to \eqref{eq:ARO_semiinfty} is certified, and the algorithm terminates with the optimal solution $(Q^{(k)}, c^{(k)})$. Otherwise, a new extreme point $y^{k}$ corresponding to the optimal solution to the subproblem \eqref{eq:AFR_subproblem} is identified. The index set $\mathcal{I}^{(k-1)}$ is updated, and a new cut is added to the master problem.





\subsubsection{Subproblem Reformulation for Improved Global Solver Performance}
It is worth noting that although the master problem \eqref{eq:AFR_master} is a tractable convex optimization problem, especially when $Q$ has low dimension, the computational bottleneck of the algorithm lies in the subproblem \eqref{eq:AFR_subproblem}, which amounts to solving an $\ell_2$-norm maximization problem over a polytope---a well-known NP-complete problem~\cite{Mangasarian1986}. Modern global solvers such as Gurobi~\cite{gurobi2024} and BARON~\cite{baron2024} handle nonconvex quadratic terms through convex relaxation and spatial branch-and-bound. Simulations show that the solver experiences extremely slow convergence with the subproblem in its original form, even with moderately-sized problems. 

To illustrate the issue, let's focus on the sum of $\ell_2$-norm term $\sum_{t=1}^T \|(D^\top y)_{[t]}\|_2$. Internally, the norm expression will be represented in terms of quadratic terms of the form $(d_i^\top y)^2$ where $d_i$ is the $i$th column of $D$. When expanded, each such term contains all cross-product terms $y_i y_k$, leading to $O(n_c^2)$ bilinear products, where $n_c$ is the row number of $D$. Summing over $T$ such norms leads to $O(n_c^2 T)$ distinct bilinear terms. Since modern global solvers build convex relaxation and perform spatial branching for each term, large branch-and-bound trees and weak relaxations ensue.

To mitigate this effect, we introduce an auxiliary vector $u = D^\top y$. The squared term $(d_i^\top y)^2$ now becomes $u_i^2$. We further introduce the auxiliary vector $w$ for the square terms ($w_i = u_i^2$). After additional auxiliary variables for the sum of $w_{[t]}$ and their square roots are introduced, the maximization of $\sum_{t=1}^T \|(D^\top y)_{[t]}\|_2$ can be reformulated as 
\begin{subequations} \label{eq:reformulation_subproblem}
\begin{align}
    \max \quad & \sum\nolimits_{t=1}^T \upsilon_t \\
    \text{s.t.} \quad & \upsilon_t^2 \le \Bigl(\frac{\sigma_t+1}{2}\Bigr)^2 - \Bigl(\frac{\sigma_t-1}{2}\Bigr)^2, && \forall t \in\mathcal{T} \\
    & \mathbbold{1}^\top w_{[t]} = \sigma_t, && \forall t \in\mathcal{T} \\
    & D^\top y = u, \\
    & \diag(u)u = w, \label{eq:norm_reform:nonconvex}
\end{align}
\end{subequations}
where the only nonconvex constraints are \eqref{eq:norm_reform:nonconvex}. The reformulation reduces the number of bilinear terms from $O(n_c^2T)$ to $O(T)$. Consequently, it yields significantly tighter convex relaxation for the norms and more efficient branch-and-bound. The computational performance improvements are corroborated in Section \ref{sect:simulation}. The reformulation of the $Q$-weighted vector norm proceeds similarly and is omitted for brevity.





\subsubsection{Stopping Criterion} The algorithm iterates between the master problem and the subproblem until the optimal value of the subproblem \eqref{eq:AFR_subproblem} becomes non-positive. Since one distinct extreme point is identified in the subproblem every iteration and the total number of extreme points is finite, the algorithm converges to the optimal solution in finite iterations. 

The overall algorithm for aggregate flexibility region characterization is summarized in Algorithm \ref{alg:fixed_topology_benders2}.

\begin{algorithm}[h]
\caption{{Aggregate flexibility characterization method}}
\label{alg:fixed_topology_benders2}
\begin{algorithmic}[1]
\State {Initialize iteration counter $k \leftarrow 0$, index set $\mathcal I^{(k)} \leftarrow \emptyset$, tolerance $\epsilon > 0$}
\State {Initialize $(Q^{(k)},c^{(k)})$}
\Repeat
    \State {$k \leftarrow k+1$}
    \State {\textbf{Solve subproblem (\ref{eq:AFR_subproblem})} to obtain the optimal value $v^{(k)}$ and the corresponding extreme point $y^k$.}
    \If{{$v^{(k)} \le \epsilon$}}
        \State {\textbf{Terminate} and return $(Q^*,c^*)=(Q^{(k)},c^{(k)})$.}
    \Else
        \State {Update $\mathcal I^{(k)} \leftarrow \mathcal I^{(k-1)} \cup \{y^k\}$.}
        \State {Add the cut (\ref{eq:AFR_master_ineqa}) to the master problem.}
    \EndIf
    \State {\textbf{Solve master problem (\ref{eq:AFR_master})} to obtain $(Q^{(k)},c^{(k)})$.}
\Until{{convergence}}
\end{algorithmic}
\end{algorithm}

%% file: section_reconfig.tex
\section{Reconfiguration for DER Aggregate Flexibility Region Optimization} \label{sect:reconfig}

Most existing works assume fixed distribution network topologies when assessing aggregate flexibility. However, modern distribution networks are reconfigurable. In this section, we formulate an optimal reconfiguration problem that optimizes the aggregate flexibility region, which builds upon the problem formulation \eqref{eq:ARO_semiinfty} and the Benders decomposition algorithm developed in Section \ref{sect:technical}. 

\subsection{Incorporation of Topology Optimization}

We denote by $\mathcal{R}$ the set of all feasible pairs $(s, \ell)$ of switch status variables and SCF auxiliary flows that satisfy the SCF radiality constraints \eqref{eq:SCF}, i.e., 
\begin{equation}
    \mathcal{R} = \bigl\{ (s, \ell) \in \{0,1\}^{|\mathcal{E}_s|} \times \mathbb{R}^{|\mathcal{E}|} \mid (s, \ell) \text{ satisfies } \eqref{eq:SCF} \bigr\}.
\end{equation}

\subsubsection{Original Formulation} 

The problem of finding the optimal topology that optimizes the ellipsoidal inner approximation of the
aggregate flexibility region can be
formulated as the following three-stage robust optimization problem:
\begin{subequations} \label{eq:ARO_reconfig}
\begin{align}
    \max_{Q \succeq 0, c} \quad & \log\det Q \label{eq:ARO_reconfig:a} \\
    \text{s.t.} \quad & \min_{(s,\ell) \in \mathcal{R}} \max_{\xi \in \mathcal{E}(Q,c),\zeta\in\mathcal{U}^T} \min_{x \in \mathcal{W}(\xi,\zeta, s)} g^\top x \le 0, \label{eq:ARO_reconfig:b} 
\end{align}
\end{subequations}
where the constraint set for the recourse variables $\mathcal{W}(\xi,\zeta, s)$ is specified in \eqref{eq:compact:Ax}. Similar to Section \ref{sect:technical}, it contains additional slack variables to ensure complete recourse. 

The formulation \eqref{eq:ARO_reconfig} searches for the largest feasible ellipsoidal flexibility region. The feasibility of the region is implied by the existence of a radial topology (the outer minimization problem in \eqref{eq:ARO_reconfig:b}) that satisfies the robust feasibility check, which is characterized by the inner ``max-min'' problem.

\subsubsection{Reformulation}
We leverage the finite-dimensional reformulation of the ``max-min'' problem in Section \ref{sect:technical:reform} to reformulate the semi-infinite problem above. Following the development therein, we obtain the reformulation below:
\begin{subequations} \label{eq:ARO_reform}
\begin{align}
    \max_{Q \succeq 0, c} \quad & \log\det Q  \\
    \text{s.t.} \quad & \min_{(s,\ell) \in \mathcal{R}} (d-Bc - Cs)^\top y^i + \norm{B^\top y^i}_Q \nonumber\\
    & + \sum\nolimits_{t=1}^T \norm{(D^\top y^i)_{[t]}}_2 \le 0, \quad i = 1, \ldots, K. \label{eq:ARO_reform:constraint}
\end{align}
\end{subequations}

{It is clear that the minimization sign in \eqref{eq:ARO_reform:constraint} can be dropped, since there exists a $(Q,c)$ so that the left-hand side is nonpositive if and only if the minimum of the left-hand side is nonpositive for some $(Q,c) \in \mathcal{R}$.} Therefore, \eqref{eq:ARO_reform} can be further simplified to arrive at the final formulation for the optimal reconfiguration problem as

\begin{mdframed}[
    skipabove=4pt,skipbelow=4pt,   
    innertopmargin=0pt,innerbottommargin=0pt,
    innerleftmargin=6pt,innerrightmargin=6pt,
    linewidth=0.7pt
]
\begingroup
\setlength{\abovedisplayskip}{0pt}
\setlength{\abovedisplayshortskip}{0pt}
\setlength{\belowdisplayskip}{3pt}
\setlength{\belowdisplayshortskip}{3pt}

\begin{subequations} \label{eq:ARO_reform2}
\begin{align}
    \max_{Q \succeq 0, c, s, \ell} \quad & \log\det Q  \\
    \text{s.t.} \quad & (d-Bc - Cs)^\top y^i + \norm{B^\top y^i}_Q \nonumber\\
    & \hspace{-0.2in} + \sum\nolimits_{t=1}^T \norm{(D^\top y^i)_{[t]}}_2 \le 0,\; i = 1, \ldots, K \label{eq:ARO_reform2:constraint}\\
    & (s,\ell) \in \mathcal{R}. \label{eq:ARO_reform2:radiality}
\end{align}
\end{subequations}

\endgroup
\end{mdframed}

Comparing the formulation above with the original one with fixed topology \eqref{eq:ARO_semiinfty:a} and \eqref{eq:bcut}, we see that the problem structures are almost identical, with the major difference being the inclusion of the SCF radiality constraints \eqref{eq:ARO_reform2:radiality}. This suggests that the problem is amenable to the same Benders decomposition algorithm described in Section \ref{sect:Benders}.

\subsection{Overall Solution Algorithm}

We now have all the components and are ready to present the full solution procedure for the optimal reconfiguration problem for aggregate flexibility region optimization. First, we discuss why the Benders decomposition algorithm developed in Section \ref{sect:Benders} remains intact despite the consideration of topological changes. The overall algorithm follows.

\subsubsection{Applicability of the Original Benders Algorithm}

At first glance, the reconfiguration problem \eqref{eq:ARO_reconfig} appears to break the Benders structure due to the three-stage framework. However, we see that the switch status variables $s$ appear in the recourse constraint \eqref{eq:compact:uncertainty} only on the right-hand side. Therefore, the dual polyhedron $\mathcal{Y}$ remains intact and is independent of the topologies. Thus, the strong-duality-based reformulation still holds. In addition, the variables $s$ enter the Benders cut \eqref{eq:ARO_reform2:constraint} only through the linear term $(Cs)^\top y^i$, which ensures that the cuts are still SOC-representable in the master problem. Finally, the minimization aligns with the signs of the inequality constraints \eqref{eq:ARO_reform:constraint} and can therefore be dropped.

\subsubsection{Overall Benders Decomposition Algorithm}
The master problem at iteration $k$ is a mixed-integer semidefinite program (MISDP) as follows:
\begin{subequations} \label{eq:reconfig_master}
\begin{align}
    \max_{Q\succeq 0, c, s, \ell} \quad & \log\det Q \\
    \text{s.t.} \quad & (d - Bc - Cs)^\top y^i + \norm{B^\top y^i}_Q \nonumber\\
    & \hspace{-0.2in} + \sum\nolimits_{t=1}^T \|(D^\top y^i)_{[t]}\|_2\le 0, \quad \forall i \in \mathcal{I}^{(k-1)} \label{eq:reconfig_master:b}\\
    & (s,\ell) \in \mathcal{R}.
\end{align}
\end{subequations}

Given a master problem solution $(Q^{(k)}, c^{(k)}, s^{(k)})$, the following subproblem is solved to verify its robust feasibility:
\begin{align} 
    \max_{y \in \mathcal{Y}}\; (d-Bc^{(k)} - Cs^{(k)})^\top y & + \norm{B^\top y}_{Q^{(k)}}  \nonumber\\
    &  + \sum\nolimits_{t=1}^T \|(D^\top y)_{[t]}\|_2. \label{eq:reconfig_subproblem}
\end{align}
When the optimal value is positive, the candidate ellipsoid is infeasible for the given topology. A new extreme point $y^k$ is identified, and a new cut in the form \eqref{eq:reconfig_master:b} is added to the master problem. On the other hand, if the optimal value is non-positive, the ellipsoid is feasible and the optimal topology $s^{(k)}$ has been identified.

The overall algorithm is summarized in Algorithm \ref{alg:afr_benders}.

\begin{algorithm}[!t]
\caption{Overall Benders decomposition algorithm for the optimal reconfiguration problem}
\label{alg:afr_benders}
\begin{algorithmic}[1]
\State Initialize iteration counter $k \gets 0$, index set $\mathcal{I}^{(k)} \gets \varnothing$, tolerance $\varepsilon>0$, $Q^{(k)}$, $c^{(k)}$, $s^{(k)}$
\Repeat
  \State $k \gets k+1$
  \State \textbf{Solve subproblem \eqref{eq:reconfig_subproblem}} with $(Q^{(k)}$, $c^{(k)}$, $s^{(k)})$ to obtain new extreme point $y^k$ and optimal cost $v^{(k)}$.
  \If{$v^{(k)} \le \varepsilon$}
    \State \textbf{Terminate} and return the optimal solution to the reconfiguration problem $(Q^*,c^*,s^*)$ as $(Q^{(k)},c^{(k)},s^{(k)})$
  \Else
    \State Update $\mathcal{I}^{(k)}\gets \mathcal{I}^{(k-1)}\cup\{y^k\}$
    \State Add cut \( (d - Bc - Cs)^\top y^i + \norm{B^\top y^i}_Q + \sum\nolimits_{t=1}^T \|(D^\top y^i)_{[t]}\|_2\le 0 \) to the master problem
  \EndIf
    \State \textbf{Solve master problem} \eqref{eq:reconfig_master} to obtain $(Q^{(k)}$, $c^{(k)}$, $s^{(k)})$
\Until{convergence}
\end{algorithmic}
\end{algorithm}

%% file: section_numerical_simulations.tex
\section{Numerical Simulations} \label{sect:simulation}

\subsection{Experiment Setup}

\subsubsection{Test Feeder Model}

In this paper, the modified IEEE 123-node test feeder \cite{bobo2021second}, \cite{kersting2001radial} with 6 controllable switches is used to verify the effectiveness of the proposed approach. Fig. \ref{fig:IEEE123} shows the position of each switch. Bus 1 is the substation bus. The load bus voltage ranges are $[0.95,1.05]\,$p.u., with the reference voltage set at $4.16$ kV. The base power is 1 MVA. The DERs include 5 HVAC units, 5 PV systems, 5 ES systems, 5 controllable loads, and 5 EVs, whose locations are shown in Table \ref{tab:der_nodes}. Loads at the remaining buses are assumed to be uncontrollable. {The HVAC parameters are summarized in Appendix~\ref{app:DER}}. The capacity of each ES system is $100$ kWh, with a charge and discharge range of $0$ to $100$ kW. Their SoCs are constrained between $0.05$ and $0.95$. The battery capacities of the EVs are assumed to follow a gamma distribution \cite{Yunfei31}, $\Gamma(4.5, 6.3)$. All EVs use level-1 (slow) charging, consistent with typical residential behavior. The maximum and minimum battery capacities are $72$ kWh and $10$ kWh, respectively. The time interval from 12:00 to 15:00 is selected for illustration purposes in this study, with a 1-hour interval. The power factor of both uncontrollable loads and DER outputs is assumed to be $0.95$. {The uncontrollable load profiles are summarized in Appendix~\ref{app:load}}. For DER parameters, $\alpha_d = 0.9$, $\beta_d \in [-0.011, -0.008]$, and $\eta_d = 0.9$ for all $d \in \mathcal{D}$. 

\subsubsection{Simulation Environment and Software}
All numerical experiments are conducted in \textsc{Matlab} R2021a on a workstation with an Intel\textsuperscript{\textregistered} Xeon\textsuperscript{\textregistered} W-1370 CPU (2.90\,GHz) and 16\,GB of RAM. 
The MISDP master problem is solved using YALMIP’s branch-and-bound \cite{Lofberg2004}, with MOSEK 10.2~\cite{MOSEK10_2} as the node solver for the continuous SDP relaxations. Gurobi 12.0~\cite{gurobi2024} is used to solve the nonconvex subproblem.

\begin{table}[!t]
    \centering
    \caption{Locations of various DERs in the test feeder}

    \caption{IEEE 123-node test feeder.}
    \label{fig:IEEE123}
\end{figure}

\subsection{Improvement of Aggregate Flexibility}
{
We set up three scenarios to compare the aggregate flexibility regions under different topologies with 5\% load uncertainty. The open switches for each scenario are listed below:
\begin{itemize}
    \item Reference topology with linear policy method (\textsc{LP-Ref}): $(13,118)$, $(54,94)$.
    \item Reference topology with Benders decomposition method (\textsc{BD-Ref}): $(13,118)$, $(54,94)$.
    \item Optimal topology with Benders decomposition method (\textsc{BD-Opt}): $(13,118)$, $(60,119)$.
\end{itemize}
}

{
The linear policy-based method is based on our previous work discussed in the literature review and is used here as a benchmark \cite{cui2021network}. The reference topology is selected as an AC-feasible radial topology and is used as a fixed-topology baseline to evaluate the benefit of topology optimization for aggregate flexibility characterization. It is assumed that the topology remains the same across all time periods.} Fig. \ref{fig:comparison_overtime} shows the power bound of each scenario. Each plot displays the flexibility region ellipsoids for adjacent periods in each scenario. {The flexibility region volume is reported in per unit, while each coordinate of the flexibility region can be converted to kW using the system base power.} The aggregate flexibility region of \textsc{LP-Ref} is 0.110, whereas that for \textsc{BD-Ref} reaches 0.264. Thus, the proposed method has a 139\% increase over the linear policy method. The linear policy imposes a linear coupling between the second-stage variables and the uncertainty, yielding a conservative feasible region than the proposed method. With topology reconfiguration enabled, \textsc{BD-Opt} achieves a volume of 0.506, representing a 91.56\% increase over \textsc{BD-Ref}. We see the aggregate flexibility region almost doubles in size after the reconfiguration, which demonstrates the effectiveness of distribution reconfiguration in unlocking DER flexibility.

\begin{figure}[!t]
    \centering
    \includegraphics[width=0.155\textwidth]{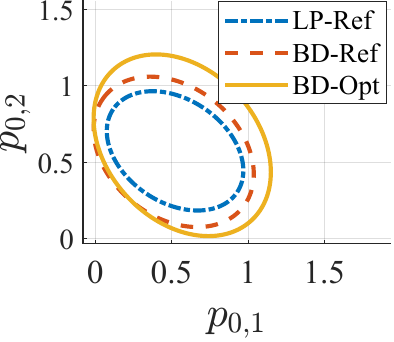}
    \includegraphics[width=0.155\textwidth]{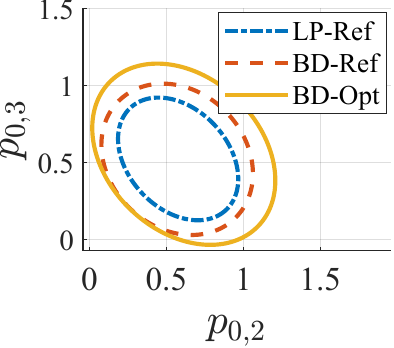}
    \includegraphics[width=0.155\textwidth]{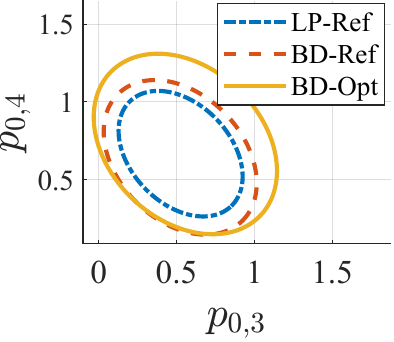}
    \caption{Flexibility region comparison {among three scenarios}.}
    \label{fig:comparison_overtime}
\end{figure}

{To further examine the performance of the proposed method, we also compare \textsc{BD-Opt} with a box-based reconfiguration benchmark inspired by~\cite{churkin2023impacts}. In this benchmark, the aggregate flexibility region is approximated by a coordinate-aligned hyperbox. For each feasible radial topology, the upper and lower bounds are obtained by solving separate optimization problems for each time period. The topology with the largest hyperbox size is selected. This benchmark is denoted as \textsc{Box-Opt}.}

\begin{table}[!t]
\caption{Comparison between Box-OPT and BD-OPT.}
\label{tab:box_benchmark3}
\centering
{
\begin{tabular}{c c c c c}
\hline
\hline
\textbf{\textsc{Method}} & \textbf{\makecell{Region \\ Type}} & \textbf{\makecell{Topology \\Treatment}} & \textbf{\textsc{Volume}} & \textbf{\textsc{Time (s)}}\\
\hline
\textsc{Box-Opt} & Hyperbox &  Enumeration & $0.0303$ & $535.8$\\
\textsc{BD-Opt} & Ellipsoid & Optimization & $0.506$ & $654.2$\\
\hline
\hline
\end{tabular}
}
\end{table}

{As shown in Table~\ref{tab:box_benchmark3}, the proposed \textsc{BD-Opt} method obtains a much larger aggregate flexibility region than the box-based reconfiguration benchmark. The result indicates that the ellipsoidal representation can better capture the directional and temporal coupling of aggregate flexibility than the coordinate-aligned hyperbox. In addition, the box-based benchmark relies on enumerating feasible radial topologies, which is tractable only because the number of candidate switches is limited, but it may become computationally expensive for systems with many controllable switches. In contrast, \textsc{BD-Opt} jointly optimizes the ellipsoidal aggregate flexibility region and the topology within the proposed Benders decomposition framework without enumerating all feasible topologies.}

\subsection{Disaggregation-Based Validation of Flexibility Region}

To further validate the benefits of network reconfiguration, a disaggregation analysis is conducted. Nine thousand disaggregation cases are simulated by randomly sampling the aggregate power trajectory $p_{0}$ and uncertainty realizations $\zeta$. With $p_{0}$ and $\zeta$ specified, a feasibility problem is formulated to determine whether there exist recourse variables that enable the system to realize the aggregate power trajectory while satisfying all operational constraints. Figure \ref{fig:voltage_comparison} reports the distribution of nodal voltages for all buses and times under \textsc{BD-Opt}. The histogram shows that all bus voltages remain feasible. This disaggregation check suggests the feasibility of the aggregate flexibility region produced by the reconfigured network. {To evaluate the impact of the LinDistFlow approximation, we further validate selected operating points using nonlinear AC power flow model. For each sampled aggregate trajectory, a feasible DER-level dispatch is recovered under the optimized topology and then tested by AC power flow model. The results show that the nodal voltages remain within the prescribed operating range and the model mismatch is small, as can be seen in Figure~\ref{fig:voltage_comparison}, which indicate that the trajectory characterized by the LinDistFlow-based model can still be delivered under the nonlinear AC model with an appropriate DER dispatch.}

In addition, certain aggregate power trajectories that are achievable under the optimal topology are infeasible under the reference topology. For example, consider the aggregate power demand vector $p_0=[0.297,1.000,0.800,0.416]$. Figure~\ref{fig:voltage_long_line} shows the voltage distribution at 13:00 under the reference topology and the optimal topology. This operating point satisfies all operational constraints under the optimal topology, while {the voltage starts to fall below the lower limit of 0.95 p.u. from bus 117 and remains below 0.95 p.u. toward the feeder end under the reference topology.} {Fig.~\ref{fig:brancch_flow} shows the branch active and reactive power flows. The optimized topology redistributes the power flow paths and alleviates heavily loaded branches compared with the reference topology. The reactive power flow is also reduced under the optimized topology. This further demonstrates that reconfiguration improves the deliverability of the aggregate flexibility by reshaping the network operating condition.} {The obtained DER-level dispatch results are shown in Fig.~\ref{fig:dispatch_results2}. The first five subplots report the dispatch of individual DERs in each category, including PV, ES, EV, HVAC, and controllable loads. Each stacked bar represents the contribution of one DER within the corresponding category at each time period. The last subplot compares the total uncertain load and the total DER dispatch. The difference between the total uncertain load and the total DER dispatch corresponds to the aggregate substation power trajectory $p_0$, which confirms that the selected aggregate trajectory can be successfully disaggregated into feasible DER-level operating points while satisfying the network constraints.}

\begin{figure}[!t]
    \centering
    \includegraphics[width=0.4\textwidth]{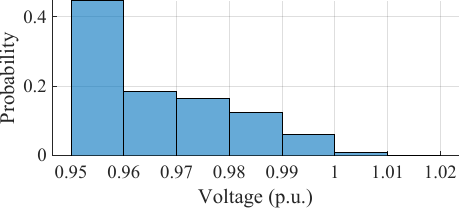}

    \includegraphics[width=0.4\textwidth]{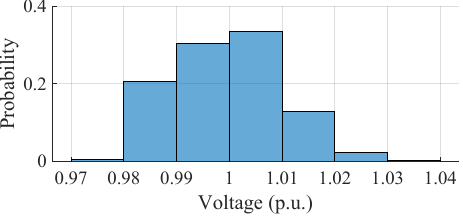}
    \caption{{Voltage distribution of all samples. (above: LinDistFlow, below: ACPF)}}
    \label{fig:voltage_comparison}
\end{figure}

    


\begin{figure}[!t]
    \centering
    \includegraphics[width=0.23\textwidth]{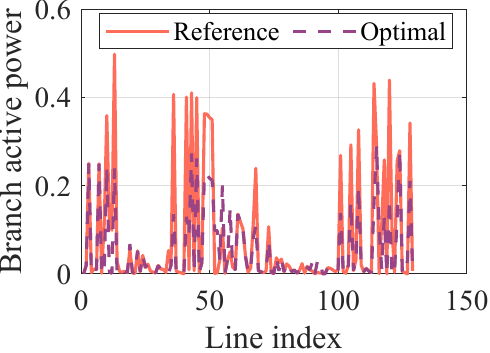}
    \includegraphics[width=0.23\textwidth]{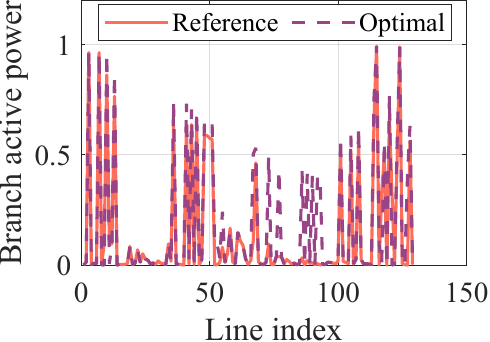}
    
    \includegraphics[width=0.23\textwidth]{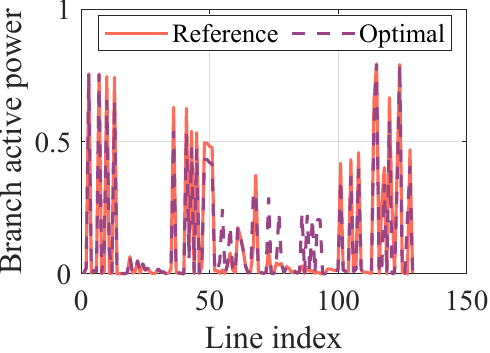}
    \includegraphics[width=0.23\textwidth]{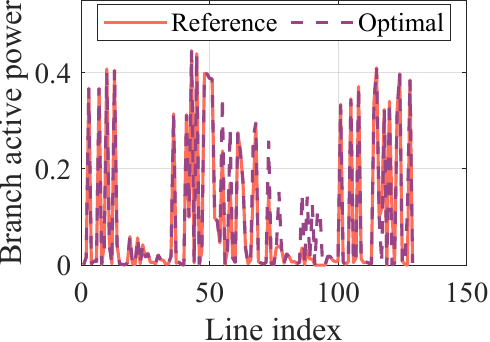}

    \includegraphics[width=0.23\textwidth]{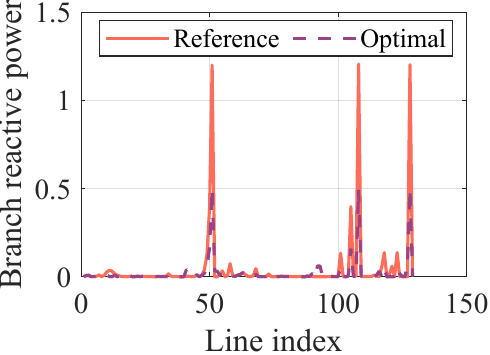}
    \includegraphics[width=0.23\textwidth]{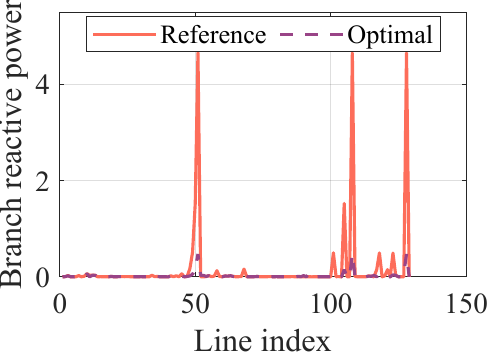}
    
    \includegraphics[width=0.23\textwidth]{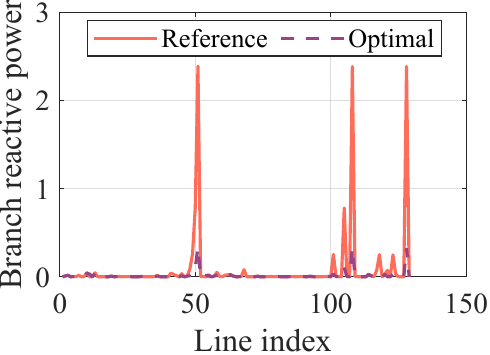}
    \includegraphics[width=0.23\textwidth]{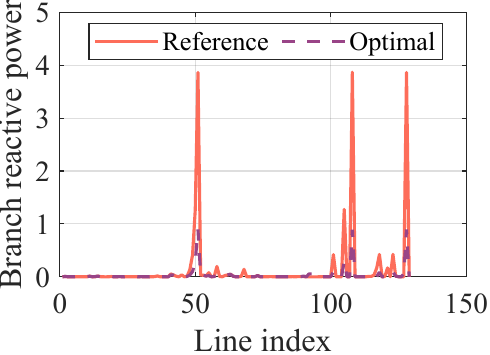}
    \caption{{Comparison of line power flow under reference and optimal topologies.}}
    \label{fig:brancch_flow}
\end{figure}

\begin{figure}[!t]
    \centering
    \includegraphics[width=0.5\textwidth]{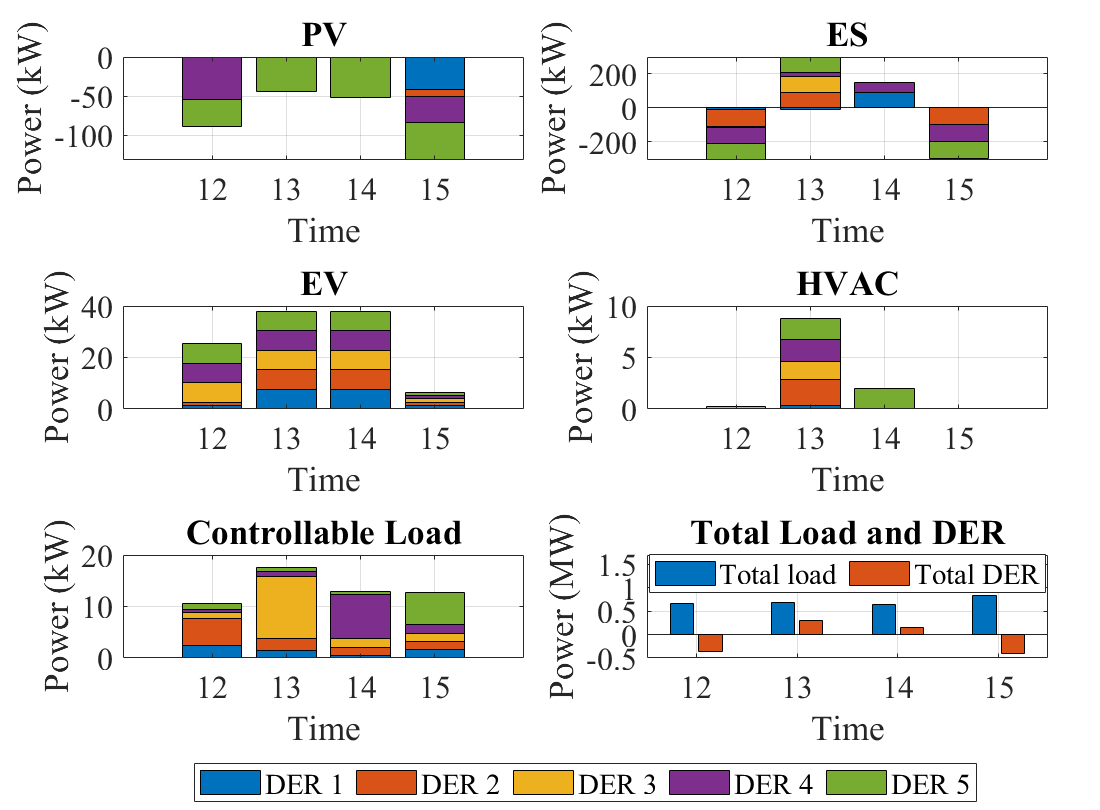}
    \caption{{Results on the aggregate flexibility disaggregation at the DER level.}}
    \label{fig:dispatch_results2}
\end{figure}

\begin{figure}[t]
    \centering

    \caption{Bus voltages at 13:00 (above: reference; below: optimal).}
    \label{fig:voltage_long_line}
\end{figure}

\subsection{Flexibility Under Different Uncertainty Levels}
The increase in uncertainty levels shrinks the aggregate flexibility region. Table \ref{tab:uncertainty level opt worst} illustrates the impact of varying uncertainty levels on flexibility volume. As uncertainty increases, the flexibility region volumes in all three scenarios decline. Despite the uniform decline, there is a clear improvement due to reconfiguration: the improvements of \textsc{BD-Opt} on \textsc{BD-Ref} range from 91.67\% to 108.3\% when the uncertainty levels lie between 5\% and 20\%. In addition, the results also demonstrate the effectiveness of the proposed aggregate flexibility region characterization method. The flexibility region characterized by the proposed method (\textsc{BD-Ref}) improves the baseline method based on our previous study (\textsc{LP-Ref}) by percentages ranging from 140.0\% to 151.2\% between 5\% and 20\% uncertainty. Overall, while the aggregate flexibility shrinks with higher uncertainty, the improvement through Benders decomposition and reconfiguration increases. Additionally, it is important to note that the optimal topology may vary under different uncertainty levels. For instance, Table \ref{tab:uncertainty level switch status} presents a comparison of the optimal switch statuses at all four uncertainty levels, with all other switches remaining closed. 


\begin{table}[!t]
    \centering
    \caption{Comparison of aggregate flexibility improvement under different uncertainty levels}
    \begin{tabular}{cccc}
        \hline
        \hline
        \textbf{\makecell{Uncertainty \\Level}} & \textbf{\textsc{LP-Ref}}  & \textbf{\textsc{BD-Ref}} & \textbf{\textsc{BD-Opt}}  \\
        \hline
        $5\%$  &    $0.110$ &  $0.264$ &   $0.506$\\
        $10\%$ &    $0.083$ &  $0.203$ &   $0.401$\\
        $15\%$ &    $0.061$ &  $0.149$ &   $0.307$\\
        $20\%$ &    $0.043$ &  $0.108$ &   $0.225$\\
        \hline
        \hline
    \end{tabular}
    \label{tab:uncertainty level opt worst}\\
\end{table}

\begin{table}[!t]
\captionsetup{font={small}}
\caption{Topology under different uncertainties}
\centering
\begin{tabular}{ccccc}
\hline
\hline
\multirow{2}{*}{\textbf{\makecell{Uncertainty \\ Level}}} & 
\multicolumn{4}{c}{\textbf{Switches Status}} \\ \cline{2-5}  
& \textbf{(13,118)} & \textbf{(60,109)} & \textbf{(54,94)} & \textbf{(117,123)} \\ \hline
5\%,10\%,20\% & 0 & 0 & 1 & 1\\ 
15\% & 1 & 1 & 0 & 0\\ 
\hline
\hline
\end{tabular}
\label{tab:uncertainty level switch status}
\end{table}

\subsection{Computation Time}
\subsubsection{Comparison of Subproblem Solution Time}
As mentioned above, Gurobi handles nonconvex quadratic terms using a spatial branch-and-bound framework built on convex relaxations. In this process, Gurobi solves convex relaxed problem, whose optimal cost lower bounds that of the original problem and is referred to as $\mathrm{Objbound}$. It also derives feasible solution for the nonconvex problem, whose optimal cost is referred to as $\mathrm{Objval}$. The search terminates when the gap between $\mathrm{Objval}$ and $\mathrm{Objbound}$ falls below the tolerance or when the time limit is reached:
\begin{subequations} \label{eq:difinition_gap}
\begin{align}
\mathrm{Gap} = \frac{|\mathrm{Objval} - \mathrm{Objbound}|}{|\mathrm{Objval}|}
\end{align}
\end{subequations}
To show the efficiency of the reformulation \eqref{eq:reformulation_subproblem}, a comparison of the computation time between the original subproblem and the reformulation suggested in \eqref{eq:reformulation_subproblem} is shown in Table \ref{tab:Comparison_subproblem_time}. The results demonstrate that the reformulation can improve computational performance. The original subproblem reaches the time limit (7200 seconds) with a gap of 24.9\textperthousand. In contrast, the reformulation is solved within 0.187 seconds, achieving a gap of 0.072\textperthousand. This confirms that the reformulation {drastically decreases the computing time and facilitates the convergence of Gurobi. Since the original subproblem already reaches the $7200\,\mathrm{s}$ time limit, it is computationally intractable to benchmark the full problem against the formulation without the subproblem reformulation.}

\begin{table}[!t]
    \centering
    \captionsetup{font={small}}
    \caption{Comparison of subproblem solution time and gap}
    \begin{tabular}{ccccc}
        \hline
        \hline
          & \textbf{Time (s)} & \textbf{\makecell{Gap (\textperthousand)}} & \textbf{Objval} & \textbf{Objbound}\\
        \hline
        \text{Original}  & $7200.8$ & $24.9$ & $-13.929$ & $-14.276$ \\
        \text{Reformulation} \eqref{eq:reformulation_subproblem} & $0.187$ & $0.072$ & $-13.929$ & $-13.930$ \\
        \hline
        \hline
    \end{tabular}
    \label{tab:Comparison_subproblem_time}
\end{table}

\subsubsection{Comparison of Total Time}
The computational efficiency of the proposed method is evaluated by measuring the time taken for each optimization. Table \ref{tab:time} presents the computing time over varying time horizons under different uncertainty levels. As maximizing aggregate flexibility region is an hours-ahead scheduling problem, even with hourly updates, the computation remains tractable for practical implementation. In this table, none of the three scenarios shows a clear trend when uncertainty levels increase. However, as the time horizon extends, the computation time increases, reflecting the higher computational demands of longer time horizons. 

Although the proposed Benders decomposition approach increases aggregate flexibility, they also lead to higher computation time. This effect becomes more evident as the time horizon increases, indicating a trade-off between achieving bigger flexibility regions and computational efficiency. 

\begin{table}[!t]
    \centering
    \captionsetup{font={small}}
    \caption{Computation time over different time spans and uncertainty levels}
    \begin{tabular}{
    >{\centering\arraybackslash}p{0.6cm}
    >{\centering\arraybackslash}p{1.1cm}
    >{\centering\arraybackslash}p{0.6cm}
    >{\centering\arraybackslash}p{1.0cm}
    >{\centering\arraybackslash}p{0.6cm}
    >{\centering\arraybackslash}p{1.0cm}
    >{\centering\arraybackslash}p{0.6cm}}
        \hline
        \hline
        \multirow{2}{*}{\textbf{\makecell{Span-\\level}}} & 
        \multicolumn{2}{c}{\textbf{\textsc{LP-Ref}}} & 
        \multicolumn{2}{c}{\textbf{\textsc{BD-Ref}}} & 
        \multicolumn{2}{c}{\textbf{\textsc{BD-Opt}}} \\
        \cline{2-7}
        & \textbf{Time(s)} & \textbf{Volume} & \textbf{Time(s)} & \textbf{Volume} & \textbf{Time(s)} & \textbf{Volume} \\
        \hline
        2-5  & $13.55$ &  $0.567$ & $28.26$ &  $0.731$ & $41.14$ &  $0.990$ \\
        2-10 & $12.59$&  $0.502$ & $23.33$&  $0.647$ & $33.77$&  $0.888$\\
        2-15 & $13.17$&  $0.441$ & $24.17$&  $0.566$ & $39.35$&  $0.790$\\ 
        2-20 & $12.50$&  $0.383$ & $28.34$&  $0.490$ & $36.52$&  $0.695$\\ \hline
        3-5  & $36.77$&  $0.281$ & $66.32$&  $0.486$ & $265.7$&  $0.765$\\
        3-10 & $36.28$&  $0.231$ & $72.79$&  $0.399$ & $114.4$&  $0.644$\\
        3-15 & $42.58$&  $0.187$ & $74.80$&  $0.323$ & $197.9$&  $0.534$\\ 
        3-20 & $38.34$&  $0.149$ & $138.2$&  $0.259$ & $257.7$&  $0.433$\\ \hline
        4-5  & $96.87$&  $0.110$ & $239.9$&  $0.264$ & $654.2$&  $0.506$\\
        4-10 & $95.83$&  $0.083$ & $374.8$&  $0.203$ & $1516$ &  $0.401$\\
        4-15 & $93.93$&  $0.061$ & $488.0$&  $0.149$ & $1455$ &  $0.307$\\
        4-20 & $95.43$&  $0.043$ & $1019$ &  $0.108$ & $2890$ &  $0.225$\\
        \hline
        \hline
    \end{tabular}
    \label{tab:time}
\end{table}

\subsection{Sensitivity to DER Penetration Levels and Candidate Switches}

{In this section, the impact of DER penetration levels and candidate switch settings on the proposed method is examined.}

\subsubsection{Impact of DER Penetration Levels}

{To examine the impact of DER penetration levels, we construct two additional cases with 50 and 75 DERs, respectively. The additional DERs are placed at randomly selected load buses while maintaining the same types of flexible resources. Table~\ref{tab:der_sensitivity3} reports the volumes of the aggregate flexibility region, number of Benders iterations, and total computation time under different DER penetration levels.}

\begin{table}[!t]
\caption{Impact of DER Penetration on Aggregate Flexibility and Computation Time}
\label{tab:der_sensitivity3}
\centering
{
\begin{tabular}{c c c c c}
\hline
\hline
\textbf{\textsc{Case}} & \textbf{\textsc{Number}} & \textbf{\textsc{Volume}} & \textbf{\textsc{Iterations }}& \textbf{\textsc{Time (s)}} \\
\hline
Base case & $25$ & $0.506$ & $29 $& $654.2 $\\
Medium DERs & $50$ & $9.245$ &$ 26$ & $283.2$ \\
High DERs & $75$ & $50.00$ & $28$ & $411.2$ \\
\hline
\hline
\end{tabular}
}
\end{table}

{The results show that increasing the number of DERs enlarges the aggregate flexibility region because more flexible resources are available for aggregation. The flexibility volume increases from $0.506$ in the base case to $9.245$ and $50.00$ in the medium- and high-DER penetration cases, respectively. The proposed Benders decomposition converges for all tested DER penetration levels, and the number of iterations remains comparable across the three cases, ranging from 26 to 29 iterations. The total computation time does not exhibit an increasing trend with the number of DERs. Although additional DERs introduce more continuous recourse variables and operational constraints, they also enlarge the feasible recourse space and provide more operational freedom for satisfying the robust feasibility conditions. As a result, the candidate flexibility region may be certified with a comparable or even smaller number of feasibility cuts. Overall, the results suggest that the proposed method remains tractable under higher DER penetration levels.}

\subsubsection{Impact of Candidate Switches}
{To examine the impact of candidate switch settings, we compare three cases: a reduced switch set, the original switch set, and an extended switch set. In the reduced switch case, switches on lines $(18,128)$, $(60,119)$, and $(97,120)$ are fixed to be closed, leaving three switches controllable. To study the impact of additional degrees of freedom from topology control, we allow three additional lines, $(44,47)$, $(76,86)$, and $(105,108)$, to be controllable in the extended switch case.}

\begin{table}[!t]
\caption{Impact of Candidate Switches on Aggregate Flexibility and Computation Time}
\label{tab:switch_sensitivity2}
\centering
{
\begin{tabular}{c c c c c}
\hline
\hline
\textbf{\textsc{Case}} & \textbf{\makecell{Switches \\ Number}} & \textbf{\textsc{Volume}} & \textbf{\textsc{Iterations}} & \textbf{\textsc{Time (s)}} \\
\hline
Reduced set & $3$ & $0.506 $& $26$ & $409.2$ \\
Original set & $6$ & $0.506$ & $29$ & $654.2$ \\
Extended set & $9 $& $0.506 $& $30$ & $527.9$ \\
\hline
\hline
\end{tabular}
}
\end{table}

{As shown in Table~\ref{tab:switch_sensitivity2}, all three candidate switch settings achieve the same flexibility volume. The reduced switch case requires less computation time than the original six-switch case, which indicates that not all candidate switches are critical for achieving the maximum aggregate flexibility region in this test case. The extended switch case also converges and achieves the same flexibility volume as the original case. This suggests that the aggregate flexibility region in this case is already limited by other binding DER or network constraints, so additional switching degrees of freedom do not further enlarge the flexibility region. The computation time does not increase monotonically with the number of candidate switches. Overall, the results show that the proposed method remains well-behaved under different candidate switch settings.}

\subsection{Scalability}
\subsubsection{Longer time horizon}

{
To further evaluate the proposed method under a longer time horizon, an additional 12-hour case study from 8:00 to 19:00 is conducted. The results in Fig.~\ref{fig:projection_area_124} show that the proposed method can still characterize multi-period aggregate flexibility regions under the 12-hour setting. Due to high PV availability during midday and afternoon periods, ES may be scheduled to absorb part of the PV output through charging to satisfy network constraints. In periods with lower PV output, ES may retain more capability to charge or discharge. Therefore, under the fixed reference topology, the region is larger in the morning hours than at other times. With network reconfiguration, the power flow paths are changed. This helps relieve voltage and branch flow constraints caused by DERs and load. Therefore, the aggregate flexibility is larger under the optimized topology, which demonstrates that network reconfiguration improves the multi-period flexibility by relieving network constraints over time. The total solution times for the optimized topology and the reference topology are $7.14$ hours and $6.78$ hours, respectively. Although the longer horizon case requires more computation time, the results indicate that the proposed framework remains applicable for extended time horizons.}

\begin{figure}[!t]
    \centering
    \includegraphics[width=0.4\textwidth]{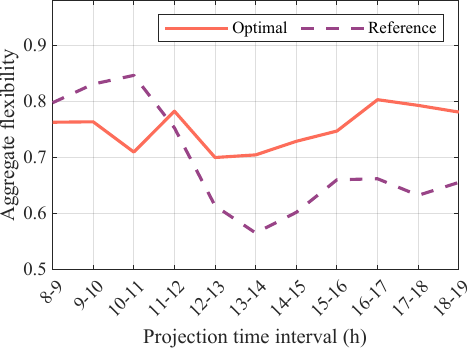}
    \caption{{The projection of aggregate flexibility for every two consecutive hours.}}
    \label{fig:projection_area_124}
\end{figure}

\subsubsection{Larger system}
{To examine the scalability, we construct a synthetic feeder by replicating the modified IEEE 123-bus feeder once and connecting the two feeder modules through additional switchable tie lines between selected buses \cite{Bahrami2024}. For comparison, the uncertainty level is fixed at $5\%$ in both cases. For the synthetic feeder with 259 nodes, the proposed method obtains a flexibility region volume of $0.856$ with a solution time of $794.824\,\mathrm{s}$. Although the larger system increases the problem size, the proposed method still solves the aggregate flexibility optimization problem within a comparable computational time. These results demonstrate that the proposed Benders decomposition-based framework is not limited to the modified IEEE 123-bus feeder and can be extended to larger distribution systems.}

%% file: section_appendices.tex
\appendices

\section{Technology-Specific DER Models} \label{app:DER}

\subsection{HVAC Units}
The power consumption of an HVAC unit $d\in\mathcal{D}$ depends on the building's temperature setpoint and the ambient temperature, which is modeled as
\begin{subequations} \label{eq:hvac}
\begin{align}
    & \ubar{p}_{d,t} \le -x_{d,t} \le \bar{p}_{d,t}, && \forall t \in \mathcal{T},  \\
    & \ubar{\theta}_{d,t} \leq \theta_{d,t} \leq \bar{\theta}_{d,t}, && \forall t \in \mathcal{T}, \\
    & \theta_{d,t} = \alpha_d \theta_{d,t}^{\mathrm{out}} + (1-\alpha_d)\theta_{d,t-1} - \beta_d x_{d,t}, && \forall t \in \mathcal{T}_2, \label{eq:5}
\end{align}
\end{subequations}
where $\ubar{p}_{d,t}$ and $\bar{p}_{d,t}$ denote the lower and upper bounds of the real power consumption at time $t$; $\theta_{d,t}$ denotes the indoor temperature at time $t$, with its lower and upper bounds denoted by $\ubar{\theta}_{d,t}$ and $\bar{\theta}_{d,t}$, respectively; $\theta_{d,t}^\mathrm{out}$ denotes the outdoor temperature at time $t$. Equation \eqref{eq:5} captures the thermal dynamics of the building. The parameters $\alpha_d$ and $\beta_d$ follow the model in \cite{Li30}. The feasible set of HVAC unit $d$ is
\begin{equation}
    \mathcal{X}_d := \bigl\{ x_d \in \mathbb{R}^T \mid \exists\, \theta_d \text{ s.t. } (x_d, \theta_d) \text{ satisfies } \eqref{eq:hvac} \bigr\}.
\end{equation}

{The HVAC power lower bounds are all zero, and the indoor temperature lower bounds are set at 68 $^\circ$F. The indoor temperature upper bounds are set to be equal to the outdoor temperature profiles, which are given as in Table~\ref{tab:hvac_outdoor_temp2}. The upper bounds of HVAC power are summarized in Table~\ref{tab:hvac_power_upper2}. }


\begin{table}[!t]
\centering
\caption{Outdoor temperature profiles for HVAC units ($^\circ$F).}
\label{tab:hvac_outdoor_temp2}
\renewcommand{\arraystretch}{1.1}
{
\begin{tabular}{cccccc}
\hline
\hline
\textbf{\textsc{Time}} & \textbf{\textsc{HVAC 1}} & \textbf{\textsc{HVAC 2}} & \textbf{\textsc{HVAC 3}} & \textbf{\textsc{HVAC 4}} & \textbf{\textsc{HVAC 5}} \\
\hline
$0$ & $82$ & $84$ & $82$ & $84$ & $84$ \\
$1$ & $86$ & $86$ & $82$ & $86$ & $86$ \\
$2$ & $90$ & $88$ & $86$ & $88$ & $88$ \\
$3$ & $91$ & $90$ & $90$ & $90$ & $90$ \\
$4$ & $91$ & $91$ & $91$ & $91$ & $90$ \\
\hline
\hline
\end{tabular}
}
\end{table}

\begin{table}[!t]
\centering
\caption{Upper bounds of HVAC power consumption (kW).}
\label{tab:hvac_power_upper2}
\renewcommand{\arraystretch}{1.1}
{
\begin{tabular}{ccccccccc}
\hline
\hline
\textbf{\textsc{Time}} & \textbf{\textsc{HVAC 1}} & \textbf{\textsc{HVAC 2}} & \textbf{\textsc{HVAC 3}}& \textbf{\textsc{HVAC 4}} & \textbf{\textsc{HVAC 5}}  \\
\hline
$1$ & $5.21$ & $0.38$ & $4.72 $ & $11.67$ & $2.10$ \\
$2$ & $6.20$ & $0.50$ & $4.56 $ & $8.10 $ & $1.76$ \\
$3$ & $6.52$ & $1.04$ & $3.44 $ & $18.82$ & $2.36$ \\
$4$ & $5.91$ & $0.41$ & $12.00$ & $9.70 $ & $1.40$ \\
\hline
\hline
\end{tabular}
}
\end{table}

\subsection{ES Systems}
An ES system provides flexibility by charging or discharging within its power and energy capacity limits. The feasible power injections of an ES system $d\in \mathcal{D}$ are constrained by:
\begin{subequations} \label{eq:ess}
\begin{align}
     & \ubar{p}_{d,t} \le x_{d,t} \le \bar{p}_{d,t}, && \forall t \in \mathcal{T},  \\
     & \ubar{e}_{d,t} \le e_{d,t} \le \bar{e}_{d,t}, && \forall t \in \mathcal{T},  \\
     & e_{d,t} = \eta_d e_{d,t-1} - \tau x_{d,t}, && \forall t \in \mathcal{T}_2,
\end{align}
\end{subequations}
where $\ubar{p}_{d,t}$ and $\bar{p}_{d,t}$ are the power limits; $e_{d,t}$ denotes the State of Charge (SoC), with bounds $\ubar{e}_{d,t}$ and $\bar{e}_{d,t}$; $\eta_{d} \in (0,1)$ is the energy retention factor. The feasible set of ES system $d$ is
\begin{equation}
    \mathcal{X}_d := \bigl\{ x_d \in \mathbb{R}^T \mid \exists\, e_{d} \text{ s.t. } (x_{d}, e_d) \text{ satisfies } \eqref{eq:ess} \bigr\}.
\end{equation}


\subsection{Controllable Loads}

A controllable load adjusts its consumption within predefined bounds. Its feasible set is 
\begin{equation}
    \mathcal{X}_d := \bigl\{ x_d \in \mathbb{R}^T \mid \ubar{p}_{d,t} \le -x_{d,t} \le \bar{p}_{d,t}, \forall t \in \mathcal{T} \bigr\}.
\end{equation}
This simple model captures the generic flexible load behavior; more complex time-coupled load models can also be incorporated if needed, but they are omitted here for simplicity.


\subsection{Electric Vehicles}
The flexibility of an EV arises from its allowable charging and its cumulative energy requirement band \cite{panda2024efficient}. The power consumption of EV $d\in\mathcal{D}$ is modeled as:
\begin{subequations} \label{eq:ev}
\begin{align}
    & \ubar{p}_{d,t} \le -x_{d,t} \le \bar{p}_{d,t}, && \forall t \in \mathcal{T},  \\ 
    & \ubar{e}_{d,t} \leq -\tau \sum_{s=1}^{t}x_{d,s} \leq \bar{e}_{d,t}, && \forall t \in \mathcal{T},
\end{align}
\end{subequations}
where $\ubar{p}_{d,t}$ and $\bar{p}_{d,t}$ denote the charging power limits; $\ubar{e}_{d,t}$ and $\bar{e}_{d,t}$ specify the cumulative energy requirement band. The feasible set of EV $d$ is 
\begin{equation}
    \mathcal{X}_d := \bigl\{ x_d \in \mathbb{R}^T \mid x_{d} \text{ satisfies } \eqref{eq:ev} \bigr\}.
\end{equation}

\subsection{PV profiles}
{The PV profiles are listed in Table~\ref{tab:pv_power_upper4}.}
\begin{table}[!t]
\centering
\caption{Upper bounds of PV available generation (kW).}
\label{tab:pv_power_upper4}
\renewcommand{\arraystretch}{1.1}
{
\begin{tabular}{cccccc}
\hline
Time & PV 1 & PV 2 & PV 3 & PV 4 & PV 5 \\
\hline
1 & 28.66 & 28.66 & 39.56 & 42.09 & 40.38 \\
2 & 36.35 & 36.35 & 46.52 & 46.37 & 57.18 \\
3 & 36.66 & 36.66 & 54.64 & 57.00 & 53.08 \\
4 & 33.84 & 33.84 & 49.41 & 51.61 & 47.71 \\
\hline
\end{tabular}
}
\end{table}

\section{Uncontrollable load profiles} \label{app:load}
\begin{figure}[!t]
    \centering
    \includegraphics[width=0.4\textwidth]{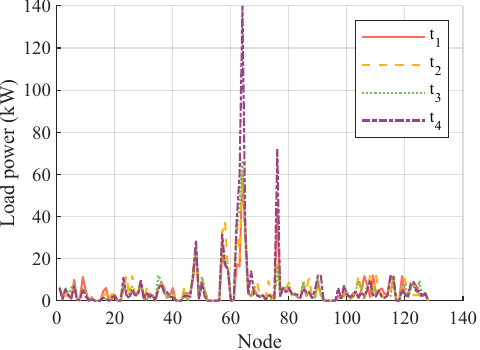}
    \caption{{Time series nodal load profile.}}
    \label{fig:time-series_nodal_load4}
\end{figure}

{The load profiles used in the case study are shown in Fig.~\ref{fig:time-series_nodal_load4}.}

%% file: IEEE_Xplore_Citation_BibTeX.bib
@ARTICLE{Wang2016decentralized,
  author={Wang, Zhaoyu and Chen, Bokan and Wang, Jianhui and Kim, Jinho},
  journal={IEEE Trans. Smart Grid}, 
  title={Decentralized Energy Management System for Networked Microgrids in Grid-Connected and Islanded Modes}, 
  month={Mar.},
  year={2016},
  volume={7},
  number={2},
  pages={1097--1105}
}

@ARTICLE{Wang2015coordinated,
  author={Wang, Zhaoyu and Chen, Bokan and Wang, Jianhui and Begovi{\'c}, Miroslav M. and Chen, Chen},
  journal={IEEE Trans. Smart Grid}, 
  title={Coordinated Energy Management of Networked Microgrids in Distribution Systems}, 
  month={Jan.},
  year={2015},
  volume={6},
  number={1},
  pages={45--53}
}

@ARTICLE{Zhao2017geometric,
  author={Zhao, Lin and Zhang, Wei and Hao, He and Kalsi, Karanjit},
  journal={IEEE Trans. Power Syst.}, 
  title={A Geometric Approach to Aggregate Flexibility Modeling of Thermostatically Controlled Loads}, 
  month={Nov.},
  year={2017},
  volume={32},
  number={6},
  pages={4721--4731}
}

@article{Alipour2022energy,
title = {Energy Storage Facilities Impact on Flexibility of Active Distribution Networks: Stochastic Approach},
journal = {Elect. Power Syst. Res.},
volume = {213},
note = {{Art.} no. 108645},
month = {Dec.},
year = {2022},
author = {Manijeh Alipour and Gevork B. Gharehpetian and Roya Ahmadiahangar and Argo Rosin and Jako Kilter},
}

@ARTICLE{Tian2022real,
  author={Tian, Guanyu and Sun, Qun Zhou and Wang, Wenyi},
  journal={IEEE Trans. Power Syst.}, 
  title={Real-Time Flexibility Quantification of a Building {HVAC} System for Peak Demand Reduction}, 
  month={Sep.},
  year={2022},
  volume={37},
  number={5},
  pages={3862--3874}
}

@article{wen2022aggregate,
  title={Aggregate feasible region of {DERs}: Exact formulation and approximate models},
  author={Wen, Yilin and Hu, Zechun and You, Shi and Duan, Xiaoyu},
  journal={IEEE Trans. Smart Grid},
  volume={13},
  number={6},
  pages={4405--4423},
  month={Nov.},
  year={2022},
  publisher={IEEE}
}

@ARTICLE{Wen2023aggregate,
  author={Wen, Yilin and Hu, Zechun and Liu, Likai},
  journal={IEEE Trans. Power Syst.}, 
  title={Aggregate Temporally Coupled Power Flexibility of {DERs} Considering Distribution System Security Constraints}, 
  month={Jul.},
  year={2023},
  volume={38},
  number={4},
  pages={3884--3896}
}

@article{wen2024improved,
  title={Improved inner approximation for aggregating power flexibility in active distribution networks and its applications},
  author={Wen, Yilin and Hu, Zechun and He, Jinhua and Guo, Yi},
  journal={IEEE Trans. Smart Grid},
  volume={15},
  number={4},
  pages={3653--3665},
  month={Jul.},
  year={2024},
  publisher={IEEE}
}

@ARTICLE{chen2020aggregate,
  author={Chen, Xin and Dall'Anese, Emiliano and Zhao, Changhong and Li, Na},
  journal={IEEE Trans. Smart Grid}, 
  title={Aggregate Power Flexibility in Unbalanced Distribution Systems}, 
  month={Jan.},
  year={2020},
  volume={11},
  number={1},
  pages={258--269}
}

@article{wang2024stochastic,
  author    = {Siyuan Wang and Wenchuan Wu and Qizhan Chen and Junjie Yu and Peng Wang},
  title     = {Stochastic flexibility evaluation for virtual power plant by aggregating distributed energy resources},
  journal   = {CSEE J. Power Energy Syst.},
  month={May},
  year={2024},
  volume={10},
  number={3},
  pages={988--999}
}

@article{wen2024centralized,
  title={Centralized distributionally robust chance-constrained dispatch of integrated transmission-distribution systems},
  author={Wen, Yilin and Hu, Zechun and Chen, Xiaolu and Bao, Zhiyuan and Liu, Chunhui},
  journal={IEEE Trans. Power Syst.},
  volume={39},
  number={2},
  pages={2947--2959},
  month={Mar.},
  year={2024},
  publisher={IEEE}
}

@ARTICLE{wang2021aggregate,
  author={Wang, Siyuan and Wu, Wenchuan},
  journal={IEEE Trans. Smart Grid}, 
  title={Aggregate Flexibility of Virtual Power Plants With Temporal Coupling Constraints}, 
  month={Nov.},
  year={2021},
  volume={12},
  number={6},
  pages={5043--5051}
}

@ARTICLE{Zhou2025aggregated,
  author={Zhou, Yihong and Essayeh, Chaimaa and Morstyn, Thomas},
  journal={IEEE Trans. Power Syst.}, 
  title={Aggregated Feasible Active Power Region for Distributed Energy Resources With a Distributionally Robust Joint Probabilistic Guarantee}, 
  month={Jan.},
  year={2025},
  volume={40},
  number={1},
  pages={556--571}
}

@article{wang2025multi,
  title={Multi-Factor-Coupled, Ahead-of-Time Aggregation of Power Flexibility Under Forecast Uncertainty},
  author={Wang, Shengyi and Du, Liang and Cui, Bai and Li, Yan},
  journal={IEEE Trans. Sustain. Energy},
  volume={40},
  number={1},
  month={Apr.},
  year={2025},
  pages={892--903},
  publisher={IEEE}
}

@ARTICLE{chen2021leveraging,
  author={Chen, Xin and Li, Na},
  journal={IEEE Trans. Smart Grid}, 
  title={Leveraging Two-Stage Adaptive Robust Optimization for Power Flexibility Aggregation}, 
  month={Sep.},
  year={2021},
  volume={12},
  number={5},
  pages={3954--3965}
}

@ARTICLE{Li2024distribution,
  author={Li, Qi and Liu, Jianzhe and Cui, Bai and Song, Wenzhan and Ye, Jin},
  journal={IEEE Trans. Smart Grid}, 
  title={Distribution System Flexibility Characterization: A Network-Informed Data-Driven Approach}, 
  month={Jan.},
  year={2024},
  volume={15},
  number={1},
  pages={1188-1191}
}

@article{cui2021network,
  title={Network-Cognizant Time-Coupled Aggregate Flexibility of Distribution Systems Under Uncertainties},
  author={Cui, Bai and Zamzam, Ahmed and Bernstein, Andrey},
  journal={IEEE Control Syst. Lett.},
  volume={5},
  number={5},
  month={Nov.},
  year={2021},
  pages={1723--1728}
}

@article{tan2024optimal,
  title={Optimal virtual battery model for aggregating storage-like resources with network constraints},
  author={Tan, Zhenfei and Yu, Ao and Zhong, Haiwang and Zhang, Xianfeng and Xia, Qing and Kang, Chongqing},
  journal={CSEE J. Power Energy Syst.},
  volume={10},
  number={4},
  month={Jul.},
  year={2024},
  pages={1843--1847},
  publisher={CSEE}
}

@article{lavorato2012imposing,
  title={Imposing radiality constraints in distribution system optimization problems},
  author={Lavorato, Marina and Franco, John F and Rider, Marcos J and Romero, Rub{\'e}n},
  journal={IEEE Trans. Power Syst.},
  volume={27},
  number={1},
  pages={172--180},
  month={Feb.},
  year={2012},
  publisher={IEEE}
}

@article{li2025aggregate,
title = {Aggregate power flexibility of multi-energy systems supported by dynamic networks},
journal = {Appl. Energy},
volume = {377},
note = {{Art.} no. 124565},
month = {Jan.},
year = {2025},
author = {Hengyi Li and Boyu Qin and Shihan Wang and Tao Ding and Jialing Liu and Hongzhen Wang}
}

@inproceedings{churkin2023impacts,
  author    = {Andrey Churkin and Miguel Sanchez-Lopez and Mohammad Iman Alizadeh and Florin Capitanescu and Eduardo A. Mart{\'i}nez Cese{\~n}a and Pierluigi Mancarella},
  title     = {Impacts of distribution network reconfiguration on aggregated {DER} flexibility},
  booktitle = {Proc. IEEE Belgrade PowerTech},
  month={Jun.},
  year      = {2023}
}

@INPROCEEDINGS{Li30,
  author={Li, Na and Chen, Lijun and Low, Steven H.},
  booktitle={Proc. IEEE Power Energy Soc. Gen. Meeting}, 
  title={Optimal demand response based on utility maximization in power networks}, 
  year={2011},
  volume={},
  number={},
  pages={1--8}}

@article{Yunfei31,
title = {A Spatial–Temporal model for grid impact analysis of plug-in electric vehicles},
journal = {Appl. Energy},
volume = {114},
pages = {456-465},
year = {2014},
author = {Yunfei Mu and Jianzhong Wu and Nick Jenkins and Hongjie Jia and Chengshan Wang}
}

@ARTICLE{Baran32,
  author={Baran, M. and Wu, F.F.},
  journal={IEEE Trans. Power Del.}, 
  title={Optimal sizing of capacitors placed on a radial distribution system}, 
  year={1989},
  month={Jan.},
  volume={4},
  number={1},
  pages={735-743}}

@article{bobo2021second,
  author  = {Lucien Bobo and Andreas Venzke and Spyros Chatzivasileiadis},
  title   = {Second-order cone relaxations of the optimal power flow for active distribution grids: Comparison of methods},
  journal = {Int. J. Electr. Power Energy Syst.},
  volume  = {127},
  note   = {{Art.} no. 106625},
  year    = {2021},
  month   = {May}
}

@inproceedings{kersting2001radial,
  author    = {William H. Kersting},
  title     = {Radial distribution test feeders},
  booktitle = {Proc. IEEE Power Eng. Soc. Winter Meeting},
  month     = {Jan.},
  year      = {2001},
  pages     = {908--912}
}

@book{bertsimas1997introduction,
  title={Introduction to Linear Optimization},
  author={Bertsimas, Dimitris and Tsitsiklis, John N},
  volume={6},
  year={1997},
  publisher={Athena Scientific},
  address={Belmont, MA, USA}
}

@article{Mangasarian1986,
author = {Mangasarian, Olvi and Shiau, T. H.},
journal = {SIAM J. Algebraic Discrete Methods},
number = {3},
pages = {455--461},
title = {A variable-complexity norm maximization problem},
volume = {7},
month = {Jul.},
year = {1986}
}

@manual{gurobi2024,
  title        = {Gurobi Optimizer Reference Manual},
  author       = {{Gurobi Optimization, LLC}},
  organization = {Gurobi Optimization, LLC},
  year         = {2024},
  note         = {Version 12.0},
  url          = {https://www.gurobi.com}
}

@manual{baron2024,
  title        = {The BARON Optimization Software},
  author       = {Sahinidis, Nikolaos V.},
  organization = {The Optimization Firm, LLC},
  year         = {2024},
  note         = {Version 24.1.13},
  url          = {https://minlp.com}
}

@article{borghetti2012mixed,
  author    = {A. Borghetti},
  title     = {A mixed-integer linear programming approach for the computation of the minimum-losses radial configuration of electrical distribution networks},
  journal   = {IEEE Trans. Power Syst.},
  volume    = {27},
  number    = {3},
  pages     = {1264--1273},
  month     = aug,
  year      = {2012},
  doi       = {10.1109/TPWRS.2012.2188913}
}

@article{ramos2005path,
  title={Path-based distribution network modeling: Application to reconfiguration for loss reduction},
  author={Ramos, E. Romero and Exp{\'o}sito, A. G{\'o}mez and Santos, Je{\'s}us Riquelme and Iborra, Francisco Llorens},
  journal={IEEE Trans. Power Syst.},
  volume={20},
  number={2},
  pages={556--564},
  month=may,
  year={2005}
}

@inproceedings{singh2022joint,
  title={Joint grid topology reconfiguration and design of {Watt}-{VAR} curves for {DERs}},
  author={Singh, Manish K and Taheri, Sina and Kekatos, Vassilis and Schneider, Kevin P and Liu, Chen-Ching},
  booktitle = {Proc. IEEE Power Energy Soc. Gen. Meeting},
  year      = {2022},
  pages     = {1--5}
}

@article{taylor2012convex,
  title={Convex models of distribution system reconfiguration},
  author={Taylor, Joshua A and Hover, Franz S},
  journal={IEEE Trans. Power Syst.},
  volume={27},
  number={3},
  pages={1407--1413},
  month=aug,
  year={2012},
  publisher={IEEE}
}

@book{boyd2004convex,
  author    = {Stephen Boyd and Lieven Vandenberghe},
  title     = {Convex Optimization},
  publisher = {Cambridge University Press},
  year      = {2004}
}

@article{panda2024efficient,
  author  = {N. K. Panda and S. H. Tindemans},
  title   = {Efficient Quantification and Representation of Aggregate Flexibility in Electric Vehicles},
  journal = {Electric Power Systems Research},
  volume  = {235},
  pages   = {110811},
  year    = {2024},
  month   = {oct}
}

@manual{MOSEK10_2,
  title  = {MOSEK Optimizer API and User Guide, Version 10.2},
  author = {{MOSEK ApS}},
  year   = {2025},
  url    = {https://docs.mosek.com}
}

@inproceedings{Lofberg2004,
author = {L{\"{o}}fberg, J.},
booktitle = {Proc. IEEE Int. Conf. Robot. Autom.},
title = {{YALMIP}: A Toolbox for Modeling and Optimization in {MATLAB}},
year = {2004},
pages = {284--289}
}

@article{dorostkar2016value,
  title={Value of distribution network reconfiguration in presence of renewable energy resources},
  author={Dorostkar-Ghamsari, Mohammad Reza and Fotuhi-Firuzabad, Mahmud and Lehtonen, Matti and Safdarian, Amir},
  journal={IEEE Trans. Power Syst.},
  volume={31},
  number={3},
  pages={1879--1888},
  month={May},
  year={2016},
  publisher={IEEE}
}

@ARTICLE{Bahrami2024,
  author={Bahrami, Shahab and Chen, Yu Christine and Wong, Vincent W. S.},
  journal={IEEE Transactions on Smart Grid}, 
  title={Dynamic Distribution Network Reconfiguration With Generation and Load Uncertainty}, 
  year={2024},
  volume={15},
  number={6},
  pages={5472-5484},
  doi={10.1109/TSG.2024.3404859}}
